\documentclass[12pt,a4paper]{article}
\usepackage{a4wide}
\usepackage{amsmath}
\usepackage{amssymb}
\usepackage{epsfig}
\usepackage{subfigure}
\usepackage{exscale}
\usepackage{float}
\usepackage{bbm}
\usepackage[numbers,sort&compress]{natbib}
\usepackage{pst-plot, pstricks,pst-math}
\usepackage{fancybox,amssymb,color}
\usepackage{graphicx}
\usepackage{pstricks, color, graphicx, epsfig, psfrag}
\usepackage{amsfonts,amsmath,amssymb,slashed}
\usepackage{dsfont}
\usepackage{bbm,bm}
\usepackage{fancyhdr, a4wide}
\usepackage[english]{babel}
\usepackage{subfigure}
\makeatletter
\@addtoreset{equation}{section}
\makeatother

\title{Emergence of Spin Order in Two-Dimensional Quantum Heisenberg Antiferromagnets}

\author{Christoph P.\ Hofmann$^a$ \\ \\
\normalsize{$^a$ Facultad de Ciencias, Universidad de Colima} \\
\vspace{0.3cm}
\normalsize{Bernal D\'iaz del Castillo 340, Colima C.P.\ 28045, Mexico} \\}

\begin{document}

\maketitle

\begin{abstract} \normalsize

Counterintuitive order-disorder phenomena emerging in antiferromagnetically coupled spin systems have been reported in various studies.
Here we perform a systematic effective field theory analysis of two-dimensional bipartite quantum Heisenberg antiferromagnets subjected to
either mutually aligned -- or mutually orthogonal -- magnetic and staggered fields. Remarkably, in the aligned configuration, the
finite-temperature uniform magnetization $M_T$ grows as temperature rises. Even more intriguing, in the orthogonal configuration, $M_T$
first drops, goes through a minimum, and then increases as temperature rises. Unmasking the effect of the magnetic field, we furthermore
demonstrate that the finite-temperature staggered magnetization $M^H_s$ and entropy density -- both exhibiting non-monotonic temperature
dependence -- are correlated. Interestingly, in the orthogonal case, $M^H_s$ presents a maximum, whereas in mutually aligned magnetic and
staggered fields, $M^H_s$ goes through a minimum. The different behavior can be traced back to the existence of an "easy XY-plane" that is
induced by the magnetic field in the orthogonal configuration.

\end{abstract}

\maketitle

\section{Introduction}
\label{Intro}

The goal of the present investigation is to achieve a more detailed understanding of the order-disorder phenomena that take place in
antiferromagnetically coupled spin systems -- foremost to assess the role of the magnetic field at finite temperature. One prominent
counterintuitive effect that has been found in many different antiferromagnetic systems is the increase of the uniform magnetization with
temperature in a magnetic field -- or, even more intriguing, an initial decrease of the uniform magnetization which presents a minimum and
only then starts to increase as temperature rises.

As far as one-dimensional systems are concerned, these rather unexpected phenomena have been reported in theoretical analyses
\citep{FKLM92,Xia98,Oku99,CEMPR02,MHO07,PHKPZP14,PR15} in a variety of settings that include isotropic, axially symmetric, easy-plane
anisotropic, and bilinear-biquadratic antiferromagnetic spin chains with different spins ($S=\frac{1}{2}, 1, \frac{3}{2}$). Experimental
studies, referring to quasi-one-dimensional Heisenberg antiferromagnets with integer spin (Haldane systems) and half-integer spin, are
Refs.~\citep{HK01,HKNH01,TSSKK05,BLIKFL13,KSAHTLT15}. The counterintuitive effect also occurs in zigzag chains and spin ladders --
theoretical investigations comprise Refs.~\citep{WY00,MO00,WOH01,YRZ04,Bouillot11,AMH14,Rez15}, while experimentally it has been found,
e.g., in the magnetic spin ladder ${(C_5 H_{12} N)}_2 Cu Br_4$ \citep{Watson01}.

Regarding two-dimensional systems, the phenomenon has been revealed in theoretical studies that tend to focus on the S=$\frac{1}{2}$
square-lattice isotropic antiferromagnet and either rely on exact diagonalization \citep{FKLM92} or Monte Carlo simulations
\citep{San99,SS02,CRVV03,PR15,Iai18}. Experimental references include the "classical" study of the quasi-two-dimensional antiferromagnet
$K_2 Mn F_4$, Ref.~\citep{AUW77}, as well as more complicated systems like anisotropic and layered antiferromagnets with different intra-
and interlayer couplings \citep{WAWLT02,Xiao09,Cizmar10,BSLPOM16,Zhao19}.

Turning to three-dimensional systems, the literature on the counterintuitive increase of the uniform magnetization with temperature is
comparatively scarce. Examples are cubic and uniaxial Heisenberg antiferromagnets \citep{Joe62}, the spin-gap magnetic compound
$Tl Cu {Cl}_3$ \citep{NOOT00}, and the $S=1$ single-ion anisotropic uniaxial antiferromagnet $Ni {Cl}_2 \! \cdot \! 4 S C {(NH_2)}_2$
\citep{PGO04}. An experimental investigation of $La_{0.17} Ca_{0.83} Mn O_3$ and  $La_{0.125} Ca_{0.875} Mn O_3$ is provided by
Ref.~\citep{BCSPDS13}. It should be pointed out that three-dimensional samples often exhibit quasi-two-dimensional behavior: the essential
physics is restricted to a plane and in the direction transverse to it the interactions are weak and hence negligible. Such materials can
hence also be described by the effective field theory results for two-dimensional systems that will be presented below.

In all these studies, except for the experimental reference \citep{AUW77}, a staggered field has not been taken into account -- the focus
rather was on the impact of the magnetic field. As is well-known, in the absence of a staggered field, antiferromagnetic systems realize
their ground state in an configuration where the staggered magnetization vector -- the order parameter -- arranges itself in a plane
perpendicular to the external magnetic field. In our analysis we also incorporate a staggered field ${\vec H_s}$, the direction of which
fixes the direction of the staggered magnetization. In physical terms, the staggered field can be interpreted as "anisotropy" field that
gives rise to the so-called easy axis along which the staggered magnetization vector aligns in a real physical sample. On top of
${\vec H_s}$, we then switch on a magnetic field $\vec H$. Here we consider two different situations: the field ${\vec H}$ either is
aligned or orthogonal to ${\vec H_s}$. The reason for this choice is that both cases have been studied in the literature although emphasis
was put on the orthogonal configuration -- in particular, a systematic effective field theory based study of the thermomagnetic properties
of antiferromagnetic monolayers in mutually aligned magnetic and staggered fields has only been undertaken very recently
\citep{Hof20a,Hof21a}.

The present theoretical investigation addresses bipartite two-dimensional quantum Heisenberg antiferromagnets subjected to magnetic and
staggered fields. Rather than following conventional microscopic approaches (modified spin-wave theory, exact diagonalization) or Monte
Carlo simulations, our systematic low-energy analysis relies on magnon effective field theory. This is the condensed matter analog of
chiral perturbation theory, i.e., pion effective field theory. The method is based on the fact that magnons, or pions, constitute the
Goldstone bosons of a spontaneously broken global symmetry. At low temperatures these are the only excited and hence relevant degrees of
freedom. If the spontaneously broken symmetry is not exact, we are dealing with pseudo-Goldstone bosons, the dispersion relations of which
are gapped. In the case of magnetic systems, the gap is due to the nonzero magnetic and staggered field -- explicit expressions are
provided below.

We show that the counterintuitive increase of the uniform magnetization with temperature generally arises in antiferromagnetic monolayers
-- irrespective of whether magnetic and staggered fields are mutually aligned or orthogonal. In the latter case we observe an even more
intriguing pattern that so far has only been reported for systems in zero staggered field (regarding two-dimensional systems, see
Refs.~\citep{FKLM92,San99,SS02,CRVV03,PR15,Iai18}): the uniform magnetization first drops, goes trough a minimum, and only then grows as
temperature rises. In the case of mutually aligned fields, on the other hand, the uniform magnetization grows monotonically with
temperature.

While the aforementioned Refs.~\citep{FKLM92,Xia98,Oku99,CEMPR02,MHO07,PHKPZP14,PR15,HK01,HKNH01,TSSKK05,BLIKFL13,KSAHTLT15,WY00,MO00,
WOH01,YRZ04,Bouillot11,AMH14,Rez15,Watson01,San99,SS02,CRVV03,AUW77,WAWLT02,Xiao09,Cizmar10,BSLPOM16,Iai18,Zhao19,Joe62,NOOT00,PGO04,
BCSPDS13} focus on the magnetic properties of the system, a discussion of the entropy -- except for Refs.~\citep{CRVV03,Joe62,BCSPDS13} --
is lacking therein. As we demonstrate in the present study, the dependence of entropy density on temperature and magnetic field strength
unambiguously reflects the order-disorder phenomena that take place in the corresponding antiferromagnetic systems. Furthermore we point
out that entropy density and finite-temperature staggered magnetization are correlated: both quantities reveal the subtle rearrangements
that occur in the antiparallel spin pattern. Here it is imperative to first unmask the effect of the magnetic field by subtracting -- both
in the entropy density and the  finite-temperature staggered magnetization -- the respective $H$=0 portions. Only then the non-monotonic
behavior of entropy density and staggered magnetization with temperature can be appreciated. As we show, non-monotonic behavior results in
either configuration: in mutually aligned or mutually orthogonal staggered and magnetic fields. But remarkably, the explicit response of
the respective system is quite different. In aligned fields, the entropy density first increases, goes through a maximum and then starts to
drop, while the finite-temperature staggered magnetization initially decreases, presents a minimum and then rises. In orthogonal fields, on
the other hand, the entropy density first drops, presents a minimum, and then starts to rise at more elevated temperatures, while the
finite-temperature staggered magnetization first rises, goes through a maximum and then falls off at more elevated temperatures.

With respect to the mechanism behind these non-monotonic and intriguing features, various explanations -- that also depend on the spatial
dimension -- have been put forward. In spin chains and ladders the phenomenon appears to be related to Luttinger liquid crossover
\citep{MHO07,WY00,WOH01,PHKPZP14}. In three-dimensional systems, Bose-Einstein condensation of magnons appears to be relevant
\citep{NOOT00,PGO04}.

In two spatial dimensions, the non-monotonic behavior of the uniform magnetization of antiferromagnetic films subjected to an external
magnetic field, has been attributed to the Kosterlitz-Thouless (KT) mechanism. The magnetic field defines an "easy plane" ("XY plane")
orthogonal to its proper direction. The spins antialign in this plane where the exchange interaction dominates and out-of-plane spin
fluctuations are suppressed -- as a result, KT behavior emerges. In particular, the minimum in the uniform magnetization -- reported in
various studies (see Refs.~\citep{FKLM92,San99,SS02,CRVV03,PR15,Iai18}) -- has been interpreted as a signature of the KT mechanism. Our
effective field theory analysis reveals that in presence of a staggered field (oriented orthogonal to the magnetic field), the minimum in
the uniform magnetization persists: the non-monotonic dependence of the uniform magnetization on temperature is also detected within the
effective field theory framework.

On the other hand, in the configuration of mutually aligned magnetic and staggered fields, there is no "easy plane". Rather, the anisotropy
field -- that is stronger than the magnetic field due to a stability criterion to be discussed below -- defines an "easy axis" along which
the spins antialign. Therefore we do not have XY (or KT) behavior. Indeed, in our effective analysis we observe simple monotonic dependence
of the uniform magnetization with temperature -- no minimum occurs here.

Regarding staggered magnetization and entropy density, as our analysis evidences, non-monotonic behavior of these quantities emerges in
either configuration of magnetic and staggered fields. In the orthogonal case where we have an XY plane and where the uniform magnetization
goes through a minimum, the finite-temperature staggered magnetization in fact presents a maximum. Naively, the magnetic field restricts
the spins to the XY plane: as such, antialignment in the plane is enforced whereas out-of-plane spin canting -- giving rise to the uniform
magnetization -- is suppressed. On the other hand, in mutually aligned magnetic and staggered fields where no easy-plane or KT behavior
emerges, the magnetic field destabilizes the antiparallel spin arrangement which leads to an increase of the uniform magnetization. In
turn, the extent of antialigned spins along the same easy axis diminishes, resulting in a decrease of the finite-temperature staggered
magnetization.

In our plots, for concreteness, we refer to the spin-$\frac{1}{2}$ square-lattice antiferromagnet -- the point is that all relevant
low-energy constants are explicitly known in this case. But we stress that the effective field theory representations of all observables
we consider here, are valid for arbitrary spin $S$ and any other bipartite two-dimensional lattice -- the only difference concerns the
concrete numerical values of spin stiffness and zero-temperature staggered magnetization (order parameter). In this perspective, the
order-disorder phenomena revealed by the non-monotonic and intriguing behavior of entropy density, finite-temperature staggered
magnetization and uniform magnetization, are universal.

The article is organized as follows. In Sec.~\ref{resultsAF} we first consider antiferromagnetic monolayers subjected to magnetic and
staggered fields that are aligned. Unmasking the impact of the magnetic field in the entropy density and the staggered magnetization,
we reveal remarkable phenomena in the thermomagnetic properties of the system that evidence destruction and creation of spin order. Along
the same lines, we then discuss the configuration of antiferromagnetic monolayers subjected to mutually orthogonal magnetic and staggered
fields, and compare the intriguing thermomagnetic effects with those occurring in the configuration of mutually aligned fields. In
Sec.~\ref{conclusions} we finally conclude. In addition, in two appendices we provide the relevant effective field theory formulae for the
entropy density, staggered and uniform magnetization for the systems underlying the present study.

\section{Thermomagnetic Properties of Antiferromagnetic Monolayers}
\label{resultsAF}

In order not to interrupt the flow of arguments, in the main body of the article we refrain from providing explicit expressions for the
uniform magnetization, staggered magnetization, and entropy density -- pertinent information is given in Appendices \ref{appendixA} and
\ref{appendixB}.

The microscopic description of antiferromagnetic monolayers is based on the quantum Heisenberg model
\begin{equation}
\label{HeisenbergZeemanH}
{\cal H} \, = \, - J \, \sum_{n.n.} {\vec S}_m \! \cdot {\vec S}_n \, - \, \sum_n {\vec S}_n \cdot {\vec H} \, - \, \sum_n (-1)^n {\vec S}_n
\! \cdot {\vec H_s} \, , \qquad J < 0 \, , \quad J = \text{const.} 
\end{equation}
Here ${\vec H}$ represents the external magnetic field and ${\vec H_s}$ stands for the staggered field. With "n.n." we indicate that the
summation is restricted to nearest neighbor spins. For concreteness, in the figures below we refer to the spin-$\frac{1}{2}$ square-lattice
antiferromagnet, but our effective field theory results are valid for any bipartite two-dimensional lattice and for arbitrary spin.

We are mainly interested in how entropy density, staggered magnetization and uniform magnetization vary with temperature and strength of
the magnetic and staggered field -- and our objective is to describe the respective order-disorder phenomena that take place in the spin
arrangement of the system. For these three observables, the general structure of the low-temperature series up to two-loop order has been
derived in Refs.~\citep{Hof17,Hof20a,Hof21a} with the outcome
\begin{eqnarray}
s(t,m,m_H) & = & s_1 T^2 + s_2 T^3 + {\cal O}(T^4) \, , \nonumber \\
M_s(t,m,m_H) & = & M_s(0,m,m_H) + {\tilde \sigma}_1 T + {\tilde \sigma}_2 T^2 + {\cal O}(T^3) \, , \nonumber \\
M(t,m,m_H) & = & M(0,m,m_H) + {\hat \sigma}_1 T + {\hat \sigma}_2 T^2 + {\cal O}(T^3) \, .
\end{eqnarray}
The respective coefficients $s_1, s_2, {\tilde \sigma}_1, {\tilde \sigma}_2, {\hat \sigma}_1, {\hat \sigma}_2$ are listed in Appendices
\ref{appendixA} and \ref{appendixB}. Note that the staggered and uniform magnetization contain a zero-temperature contribution:
$M_s(0,m,m_H)$ and $M(0,m,m_H)$, respectively. Instead of working with absolute values of field strengths $H_s, H$, and temperature $T$, we
prefer to use the dimensionless parameters
\begin{equation}
\label{definitionRatios}
m \equiv \frac{\sqrt{M_s H_s}}{2 \pi \rho_s^{3/2}} \, , \qquad
m_H \equiv \frac{H}{2 \pi \rho_s} \, \, , \qquad
t \equiv \frac{T}{2 \pi \rho_s} \, .
\end{equation}
The motivation for these definitions is that the common denominator,
\begin{equation}
2 \pi \rho_s \approx J \, ,
\end{equation}
is of the order of the exchange coupling $J$ that defines the microscopic scale. In the domain where the low-energy effective field theory
is valid, the parameters $m, m_H, t$ are small. In subsequent plots we go up to
\begin{equation}
\label{domain}
m, m_H, t \ \lessapprox 0.3 \, .
\end{equation}

\subsection{Antiferromagnetic Monolayers in Mutually Parallel Magnetic and Staggered Fields}
\label{caseI}

We first address the configuration where magnetic and staggered fields are mutually parallel:
\begin{equation}
\label{externalFields}
{\vec H} = (H,0,0) \, , \qquad {\vec H}_s = (H_s,0,0) \, , \qquad H, H_s > 0 \, .
\end{equation}
Note that ${\vec H}$ and ${\vec H}_s$ point into the direction of the order parameter (staggered magnetization at $T$=0). In presence of
these fields, the dispersion laws for the two magnons take the form,\footnote{The spin-wave velocity $v$ we have set to one.}
\begin{eqnarray}
\label{disprelAFHparallel}
\omega_{+} & = & \sqrt{{\vec k \,}^2 + \frac{M_s H_s}{\rho_s}} + H \, , \nonumber \\
\omega_{-} & = & \sqrt{{\vec k \,}^2 + \frac{M_s H_s}{\rho_s}} - H \, ,
\end{eqnarray}
where $M_s$ is the staggered magnetization at zero temperature (and zero magnetic and staggered field) and $\rho_s$ is the spin stiffness.

It should be mentioned that $\omega_{-}$ becomes negative, unless the stability criterion
\begin{equation}
\label{stabilityCondition}
H_s > \frac{\rho_s}{M_s} \, H^2
\end{equation}
is satisfied. Here we assume this is indeed the case. More concretely, in the plots we will restrict ourselves to the parameter domain
\begin{equation}
\label{stabilityCriterion}
m > m_H + \delta \, , \qquad  \delta = 0.03 \, .
\end{equation}
If the stability condition is not met, the direction of the staggered magnetization vector changes: the system -- in a so-called spin-flop
transition -- evolves into a configuration where staggered magnetization and external magnetic field are oriented orthogonal. This
situation will be analyzed in subsection \ref{caseII}.

\begin{figure}
\begin{center}
\hbox{
\includegraphics[width=7.2cm]{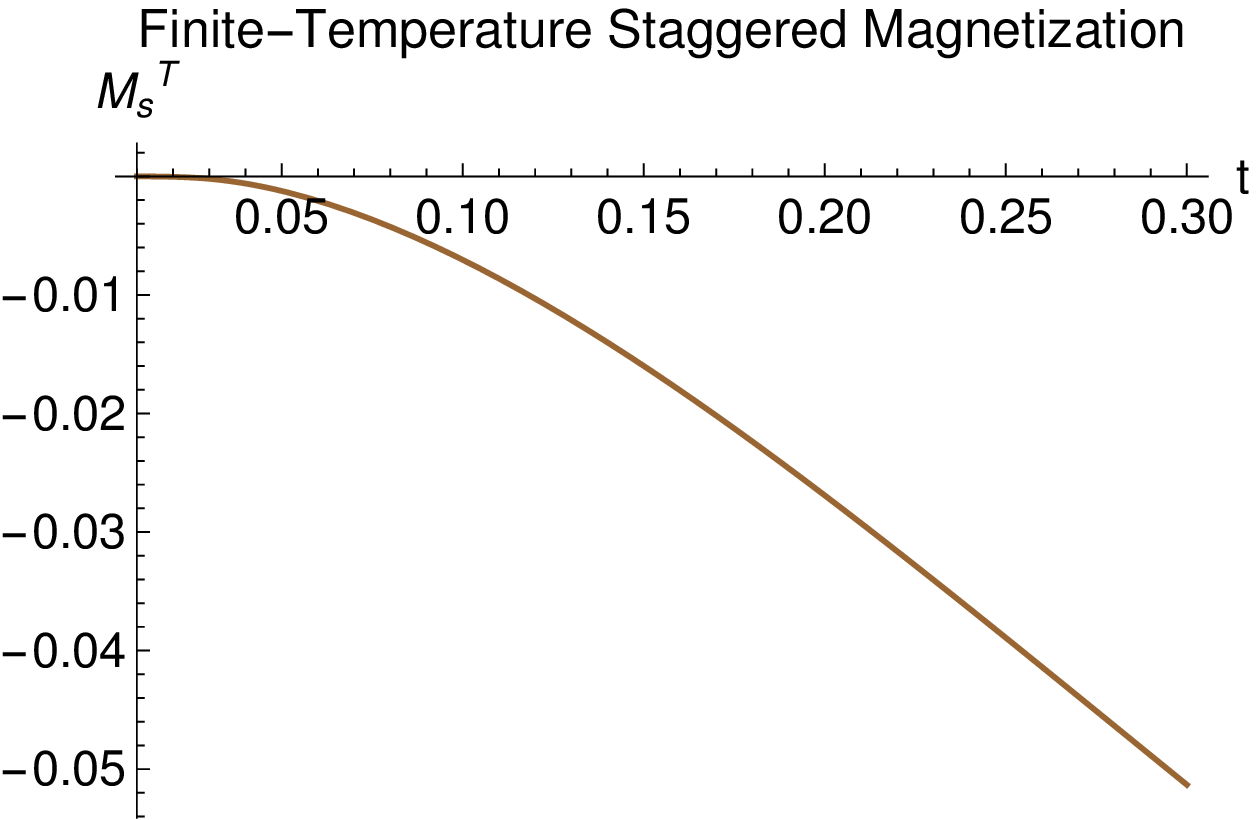}
\hspace{2mm}
\includegraphics[width=7.2cm]{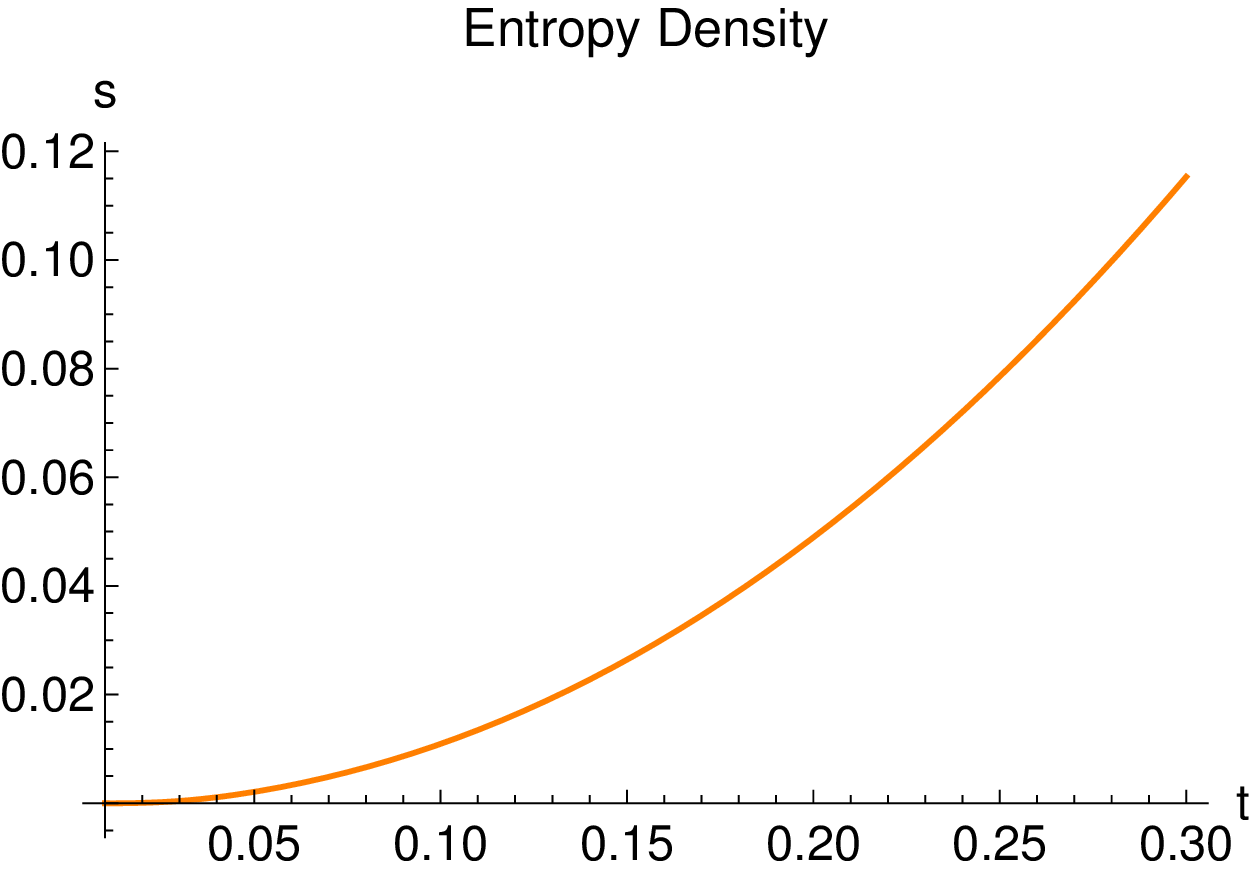}}
\end{center}
\caption{[Color online] Temperature dependence of the staggered magnetization $M^T_s$ and entropy density $s$ for the spin-$\frac{1}{2}$
square-lattice antiferromagnet in mutually parallel staggered and magnetic fields of strength $(m,m_H)=(0.25,0.15)$.}
\label{figure1}
\end{figure}

To gain a rough idea on the thermomagnetic behavior of antiferromagnetic monolayers in mutually parallel magnetic and staggered fields,
in Fig.~\ref{figure1} we show the dominant feature: as temperature rises, the finite-temperature staggered magnetization $M^T_s$ decreases,
while the entropy density $s$ increases. In the plot, staggered and magnetic field strengths are held fixed: concretely we refer to the
point $(m,m_H)=(0.25,0.15)$. Note that the finite-temperature staggered magnetization is defined as
\begin{equation}
\label{MTs}
M^T_s(t,m,m_H) = M_s(t,m,m_H) - M_s(0,m,m_H) \, ,
\end{equation}
i.e., the $T$=0 contribution in the total staggered magnetization has been subtracted. The finite-temperature portion $M^T_s$ therefore
measures the change of the staggered magnetization when temperature is raised from $t$=0 to $t \neq 0$. The above finding is not really
spectacular -- after all, this is what one would expect intuitively: thermal fluctuations destabilize the antiferromagnetic spin order
and, as a consequence, the entropy grows. But interesting effects show up when one considers the impact of the magnetic field alone.

\begin{figure}
\begin{center}
\hbox{
\includegraphics[width=7.2cm]{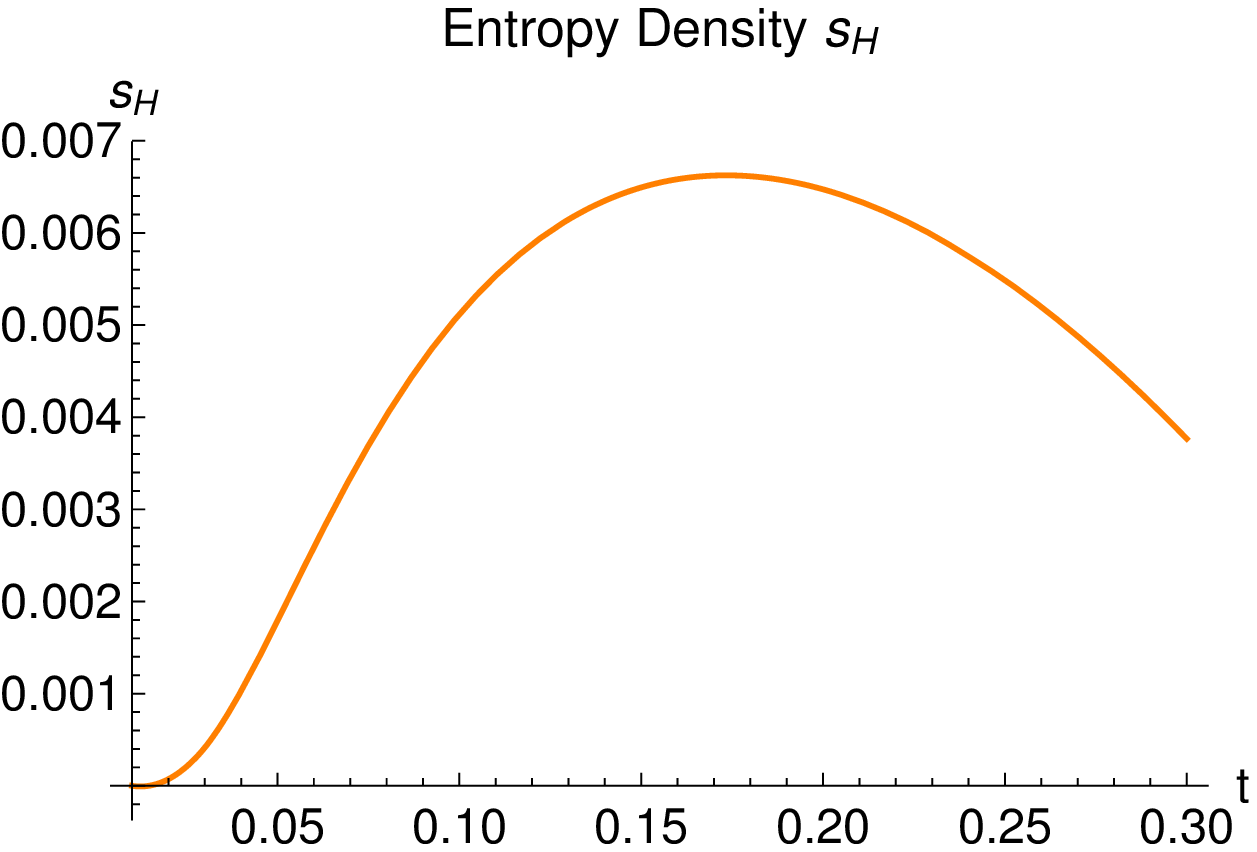}
\hspace{2mm}
\includegraphics[width=7.2cm]{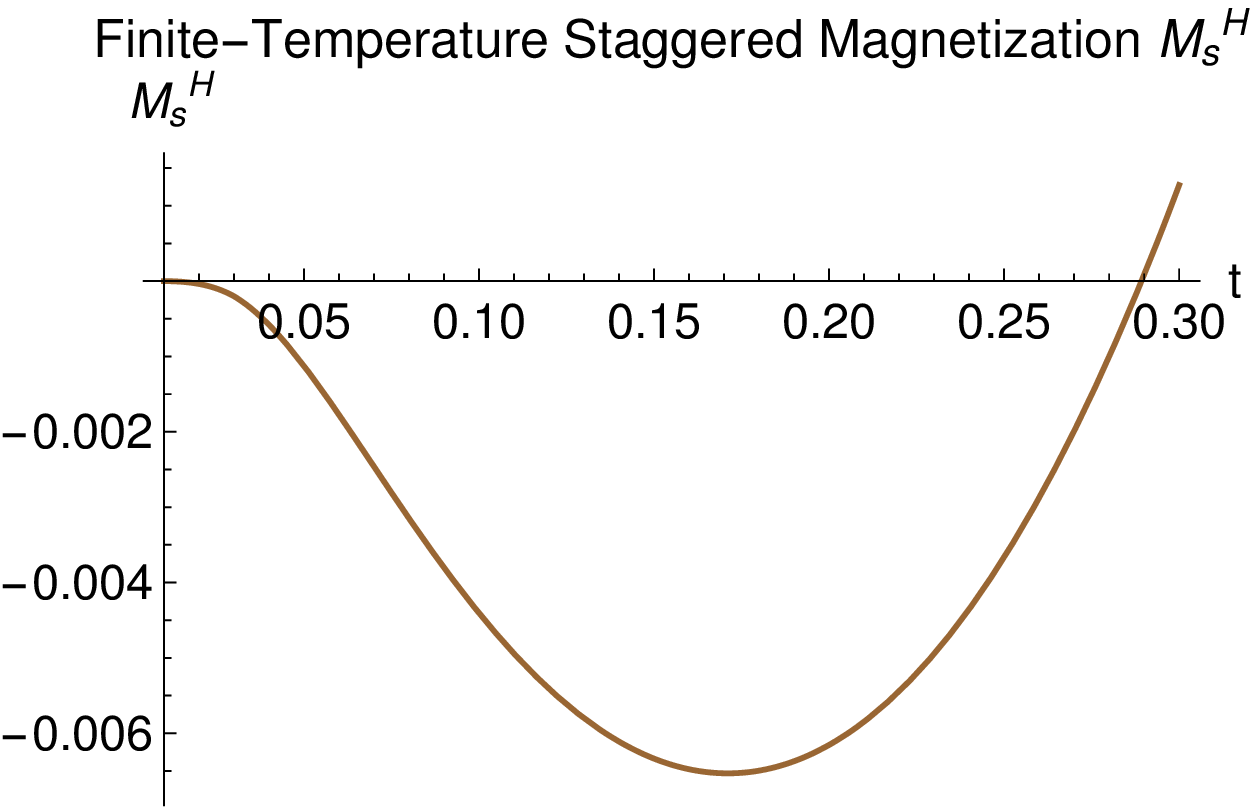}}
\vspace{4mm}
\includegraphics[width=7.2cm]{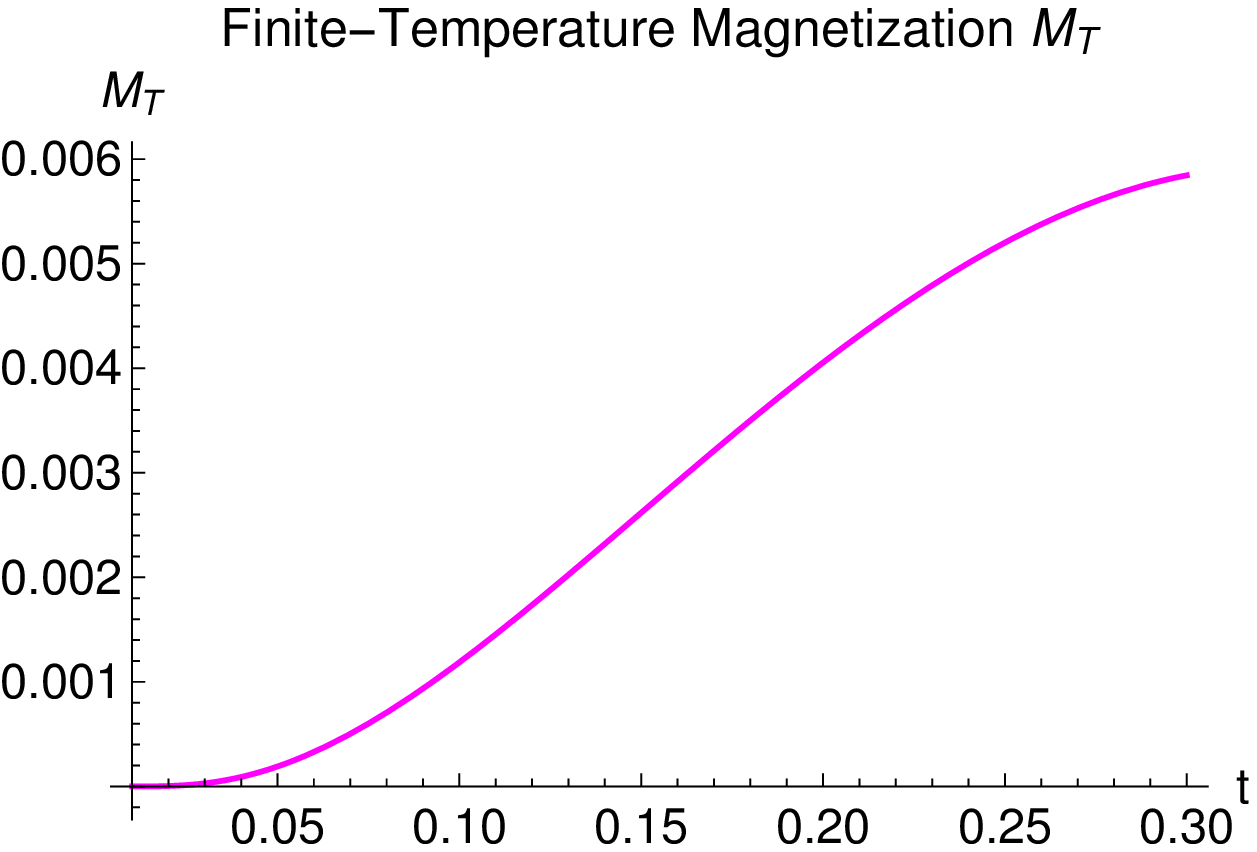}
\end{center}
\caption{[Color online] Temperature dependence of the $H$-induced entropy density $s_H$, staggered magnetization $M^H_s$, and uniform
magnetization $M_T$ of the spin-$\frac{1}{2}$ square-lattice antiferromagnet in mutually parallel staggered and magnetic fields of strength
$(m,m_H)=(0.25,0.15)$.}
\label{figure2}
\end{figure}

To unmask the effect of the magnetic field, in the finite-temperature staggered magnetization and entropy density, we now subtract the
portions that are uniquely due to the staggered field,
\begin{equation}
M^H_s = M_s^T(t,m,m_H) - M_s(t,m,0)  \, , \qquad s_H = s(t,m,m_H) - s(t,m,0) \, .
\end{equation}
The quantities $M^H_s$ and $s_H$ monitor the changes of the staggered magnetization and entropy density with temperature that are uniquely
caused by the magnetic field. Remarkably, as illustrated in the upper panel of Fig.~\ref{figure2}, $M^H_s$ and $s_H$ are correlated and --
unlike $M^T_s$ and $s$ depicted in Fig.~\ref{figure1} -- exhibit non-monotonic behavior. The entropy density first increases, goes through
a maximum and then starts to drop, while $M^H_s$ behaves in the opposite way: it initially decreases, presents a minimum and then rises.
Note that the entropy maximum at $t^{\text{max}}_s=0.173$ and the staggered magnetization minimum at $t^{\text{min}}_{M_s}=0.171$ almost
coincide. The quantities $M^H_s$ and $s_H$ witness the counterintuitive phenomenon that, in presence of a magnetic field, antiparallel spin
order is initially destroyed at low temperatures, but subsequently reestablished at more elevated temperatures.

Furthermore, as shown in the lower panel of Fig.~\ref{figure2}, in presence of a magnetic field, a finite-temperature uniform magnetization
$M_T$,\footnote{As in the staggered magnetization, Eq.~(\ref{MTs}), we subtract the $T$=0 portion from the total uniform magnetization.
Accordingly, $M_T$ indicates how the uniform magnetization changes when temperature is raised from $t$=0 to $t \neq 0$.}
\begin{equation}
M_T = M(t,m,m_H)- M(0,m,m_H) \, ,
\end{equation}
is induced along the staggered magnetization axis. Remarkably, $M_T$ increases as temperature rises -- quite the opposite of what one would
expect intuitively. Still, up to the temperature $t^{\text{min}}_{M_s}=0.171$, the dominant effect is the destruction of antiferromagnetic
spin alignment, witnessed by the finite-temperature staggered magnetization $M^H_s$ and reflected in the entropy density $s_H$. Most
importantly, the correlation is between entropy density and finite-temperature staggered -- and not uniform -- magnetization.

\begin{figure}
\begin{center}
\includegraphics[width=7.5cm]{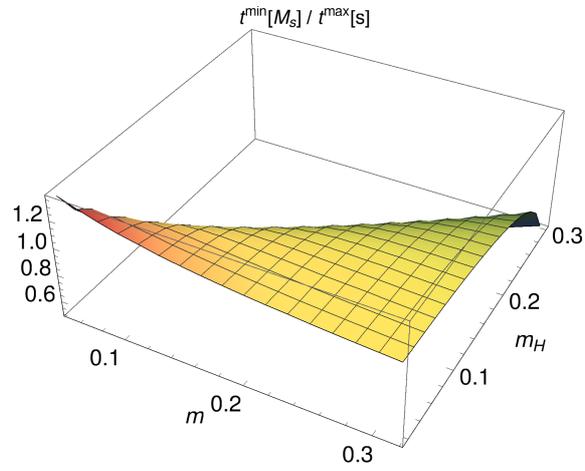}
\end{center}
\caption{[Color online] Correlation between staggered magnetization minimum and entropy density maximum for the spin-$\frac{1}{2}$
square-lattice antiferromagnet subjected to mutually parallel magnetic ($m_H$) and staggered fields ($m$).}
\label{figure3}
\end{figure}

This correlation not only occurs for the specific point $(m,m_H)=(0.25,0.15)$, but can be observed in the entire parameter region as we
illustrate in Fig.~\ref{figure3}. Larger deviations of the ratio $t^{\text{min}}_{M_s}/t^{\text{max}}_{s}$ from $1$ only start showing up in
stronger magnetic fields where the effective expansion is about to break down, or in staggered fields not much larger than the magnetic
field, where we approach the regime where the stability criterion (\ref{stabilityCriterion}) no longer is satisfied. Overall, the picture
is consistent: the initial destruction of antiferromagnetic spin order caused by the magnetic field is manifested simultaneously in the
decrease of the finite-temperature staggered magnetization $M^H_s$ and in the increase of the entropy density $s_H$.

\begin{figure}
\begin{center}
\hbox{
\includegraphics[width=6.5cm]{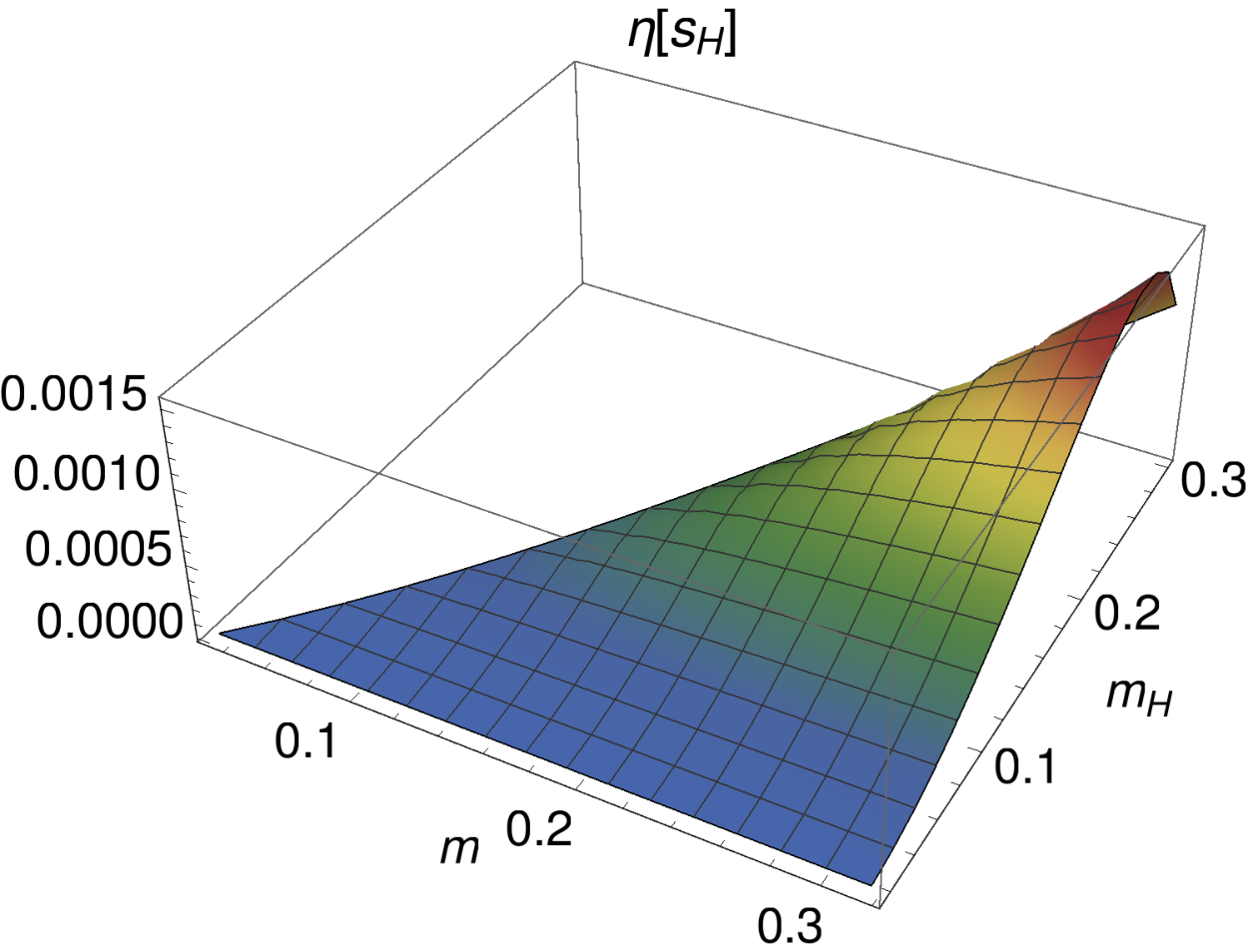}
\hspace{2mm}
\includegraphics[width=6.5cm]{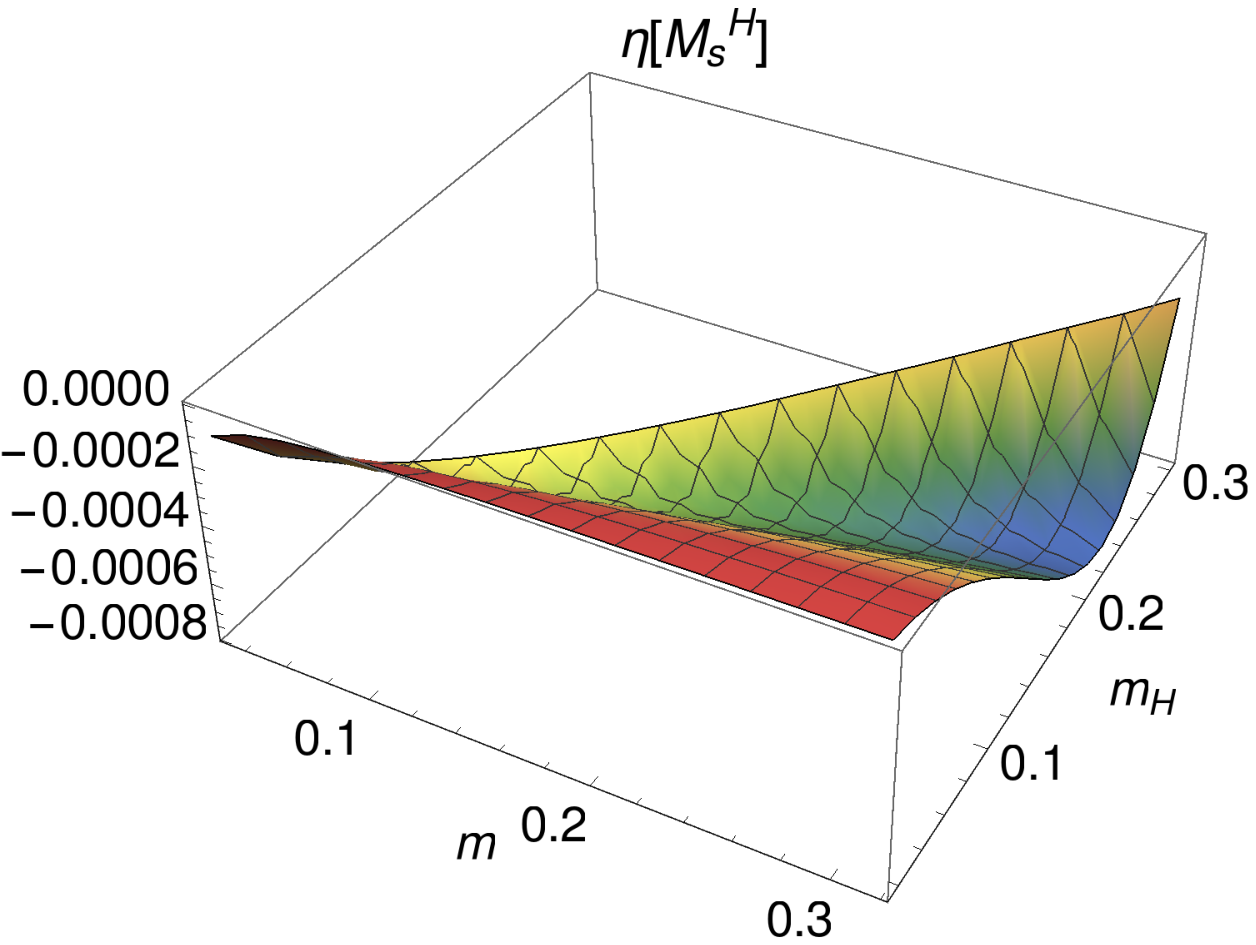}}
\end{center}
\caption{[Color online] The parameters $\eta[s_H]$ and $\eta[M^H_s]$ for the spin-$\frac{1}{2}$ square-lattice antiferromagnet subjected to
mutually parallel magnetic ($m_H$) and staggered ($m$) fields.}
\label{figure4}
\end{figure}

To put these observations on more quantitative grounds, we now consider the area under the entropy density curve (see Fig.~\ref{figure2})
between the temperatures $t$=0 and the maximum at $t^{\text{max}}_s$,
\begin{equation}
\eta[s_H] = \int_0^{t^{\text{max}}_{s}} \!\!\! dt \, s_H \, .
\end{equation}
The parameter $\eta[s_H]$ measures the initial increase of entropy density caused by the magnetic field. Analogously, for the
finite-temperature staggered magnetization, we define the parameter
\begin{equation}
\eta[M^H_s] = \int_0^{t^{\text{min}}_{M_s}} \!\!\! dt \, M^H_s
\end{equation}
that measures the initial destruction of antiferromagnetic alignment caused by the magnetic field. To capture the thermomagnetic properties
of the system in the entire parameter space defined by magnetic and staggered field strength, we scan the surface $(m,m_H)$ and evaluate
$\eta[s_H]$ and $\eta[M^H_s]$ for each selected point.\footnote{It should be noted that the extrema $t^{\text{max}}[s]$ and $t^{\text{min}}[M_s]$
for any points of our scan $(m,m_H)$ lie within the temperature interval $0 < t \lessapprox 0.3$ where the low-energy effective field
theory applies.} The result is shown in Fig.~\ref{figure4}. The destabilization of antiferromagnetic order by the magnetic field can be
observed simultaneously in the entropy density and the staggered magnetization. The perturbation of the antialigned spins gets stronger as
the magnetic field strength grows, but eventually the effect is damped.

\begin{figure}
\begin{center}
\includegraphics[width=7.5cm]{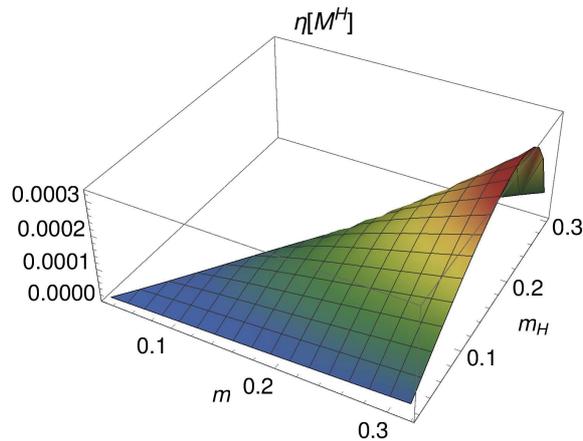}
\end{center}
\caption{[Color online] The parameter $\eta[M^H]$ for the spin-$\frac{1}{2}$ square-lattice antiferromagnet subjected to mutually parallel
magnetic ($m_H$) and staggered ($m$) fields.}
\label{figure5}
\end{figure}

To underline that the destruction of antiparallel spin alignment is the dominant effect -- and not the creation of a uniform magnetization
along the order parameter axis -- we define the parameter $\eta[M^H]$, associated with the uniform magnetization, as
\begin{equation}
\eta[M^H] = \int_{0}^{t^{\text{min}}_{M_s}} \!\!\! dt \, M_T \, .
\end{equation}
To be able to compare this parameter with $\eta[M^H_s]$ and $\eta[s_H]$, it is important to integrate the uniform magnetization curve up to
the same temperature as before that is defined by the entropy maximum (or, equivalently, by the staggered magnetization
minimum).\footnote{In fact, the uniform magnetization maximum for the specific point $(m,m_H)=(0.25,0.15)$ -- as for most other points in
parameter space $(m,m_H)$ -- is located at $t > 0.3$, i.e., outside the low-temperature domain where the effective field theory applies.}
According to Fig.~\ref{figure5}, the creation of a uniform magnetization becomes more pronounced as the magnetic field strength grows, but
eventually the effect is damped in stronger magnetic fields. However notice that in the entire domain $(m,m_H)$, the parameter $\eta[M^H]$
is smaller than $\eta[M^H_s]$ depicted in Fig.~\ref{figure4}: the creation of a uniform magnetization is not the dominant effect -- what
counts in the entropy density $s_H$ is the destruction of antiferromagnetic order.

Although the configuration of mutually aligned magnetic and staggered fields is motivated from a physical point of view and is even
discussed in reviews and textbooks (see, e.g., Refs.~\citep{ABK61a,Nol86}), remarkably, the counterintuitive thermomagnetic properties of
this system -- except for the effective field theory based Refs.~\citep{Hof20a,Hof21a} -- have not been studied so far analytically or by
Monte Carlo simulations. Regarding the experimental side, we are only aware of the "classical" Ref.~\citep{AUW77} that focuses on the
thermomagnetic behavior of the specific quasi two-dimensional antiferromagnet $K_2 Mn F_4$. The emergence of a uniform magnetization that
grows with temperature observed in this sample is consistent with what we find.

\subsection{Antiferromagnetic Monolayers in Mutually Orthogonal Magnetic and Staggered Fields}
\label{caseII}

Let us now address the configuration of mutually orthogonal fields,\footnote{The staggered field ${\vec H}_s$ fixes the direction of the
order parameter while the magnetic field ${\vec H}$ lies in a plane transverse to the order parameter.}
\begin{equation}
{\vec H} = (0,H,0) \, , \qquad {\vec H}_s = (H_s,0,0) \, , \qquad H, H_s > 0 \, .
\end{equation}
In this case the two magnons obey the dispersion relations
\begin{eqnarray}
\label{disprelAFH}
\omega_{I} & = & \sqrt{{\vec k}^2 + \frac{M_s H_s}{\rho_s} + H^2} \, , \nonumber \\
\omega_{I\!I} & = & \sqrt{{\vec k}^2 + \frac{M_s H_s}{\rho_s}} \, .
\end{eqnarray}
Notice that the dispersion law of magnon $I\!I$ is not affected by the magnetic field.

\begin{figure}
\begin{center}
\hbox{
\includegraphics[width=7.2cm]{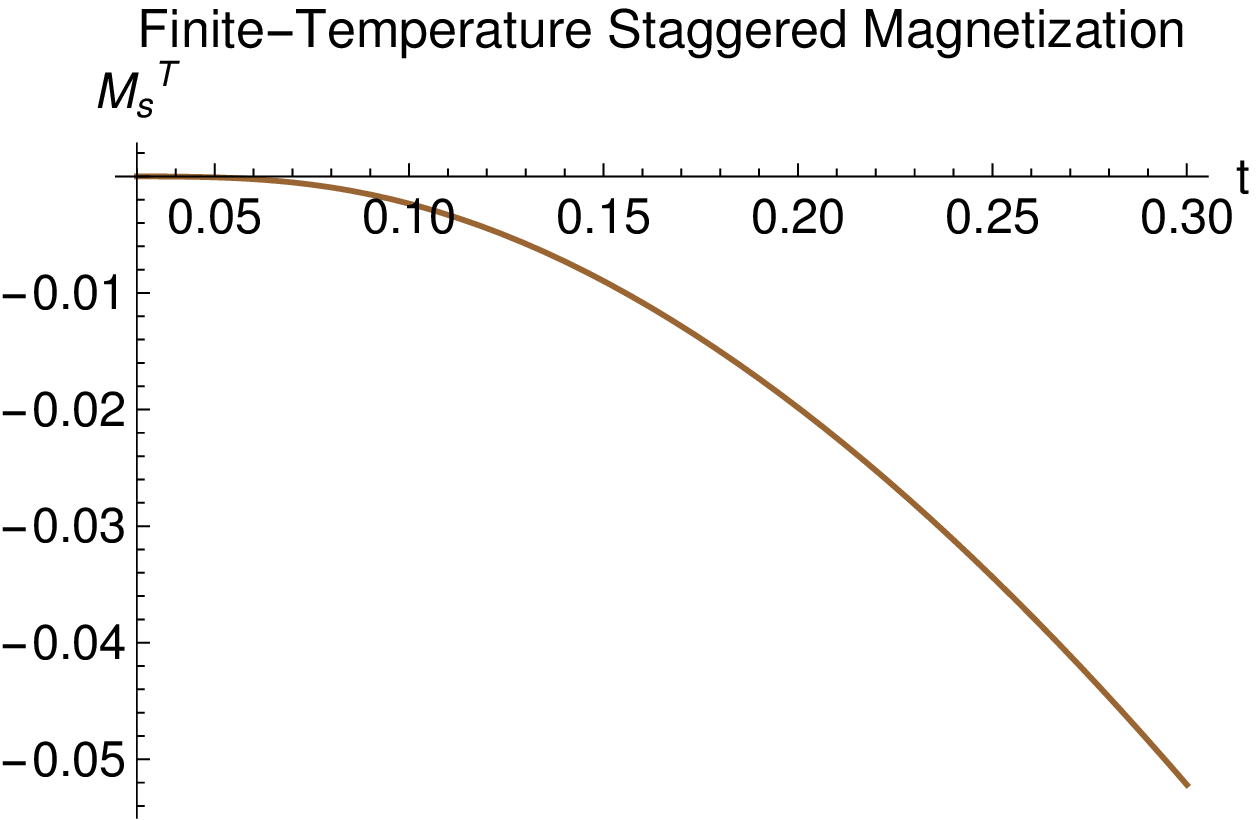}
\hspace{2mm}
\includegraphics[width=7.2cm]{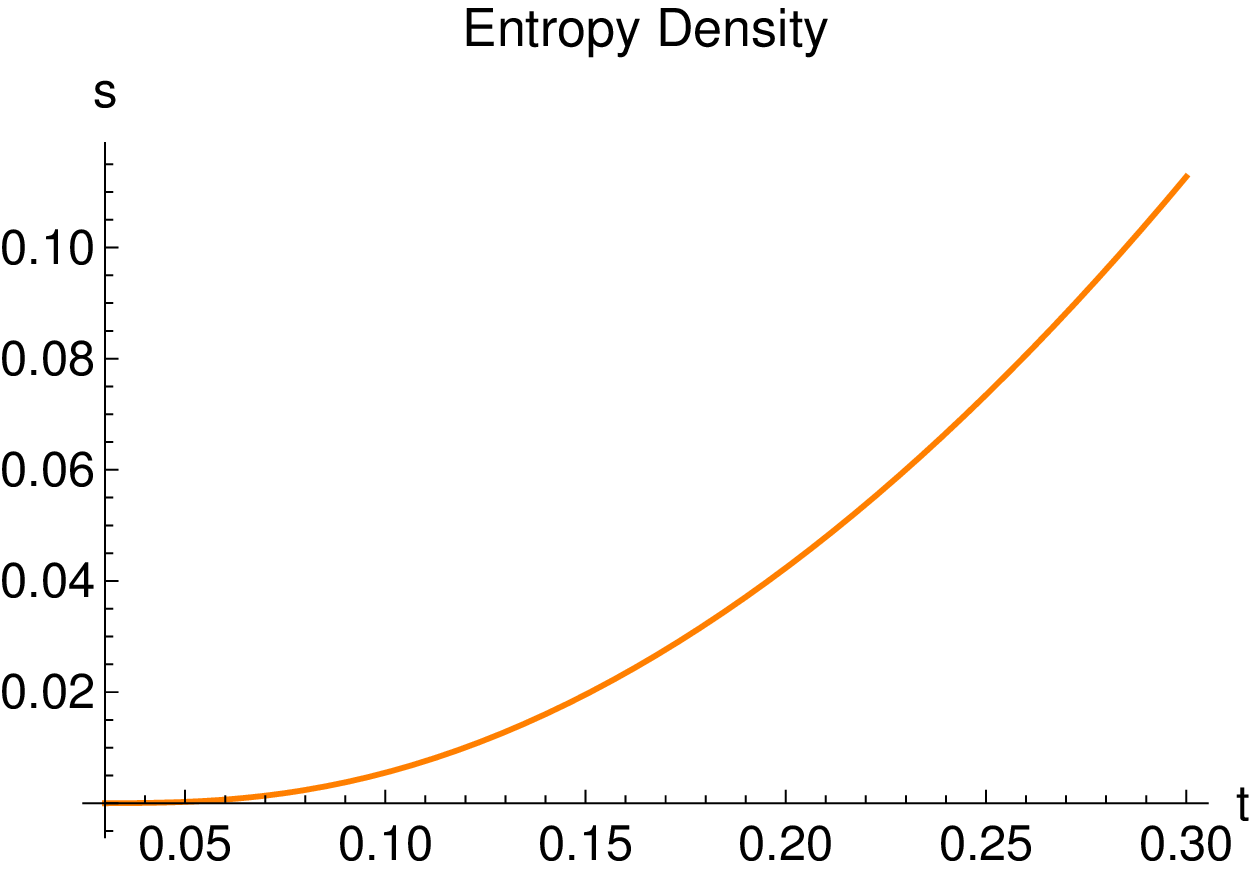}}
\end{center}
\caption{[Color online] Temperature dependence of staggered magnetization $M^T_s$ and entropy density $s$ for the spin-$\frac{1}{2}$
square-lattice antiferromagnet in mutually orthogonal staggered and magnetic fields of strength $(m,m_H)=(0.25,0.15)$.}
\label{figure6}
\end{figure}

In the configuration of mutually orthogonal fields, no stability criterion must be met, but we have to keep in mind that the effective
finite-temperature field theory framework starts to break down in very weak staggered fields -- in particular, the limit $H_s \to 0$ cannot
be taken. This is a consequence of the Mermin-Wagner theorem and has been discussed previously, e.g., in Sec.~V of Ref.~\citep{Hof10}, and
also illustrated by Figs.~2 and 3 of Ref.~\citep{Hof16b}. The same caveat in fact also applies to the case of mutually aligned fields.
However, if the stability criterion is met, the restriction imposed by the Mermin-Wagner theorem is automatically satisfied. The crucial
point is that all plots presented here refer to parameter regions where our effective field theory results are perfectly valid.

To get a rough image of the thermomagnetic behavior of the system, in Fig.~\ref{figure6} we show the dominant characteristic: as
temperature rises, the finite-temperature staggered magnetization $M^T_s$ -- defined in Eq.~(\ref{MTs}) -- decreases, while the entropy
density $s$ increases. In the plots, where staggered and magnetic field strengths are held fixed, we have chosen the same point
$(m,m_H)=(0.25,0.15)$ as in Fig.~\ref{figure1} and also use the same microscopic units. The effect is qualitatively and quantitatively the
same as in the case of mutually parallel fields and the intuitive picture is confirmed: thermal fluctuations destabilize the
antiferromagnetically ordered spins.

\begin{figure}
\begin{center}
\hbox{
\includegraphics[width=7.2cm]{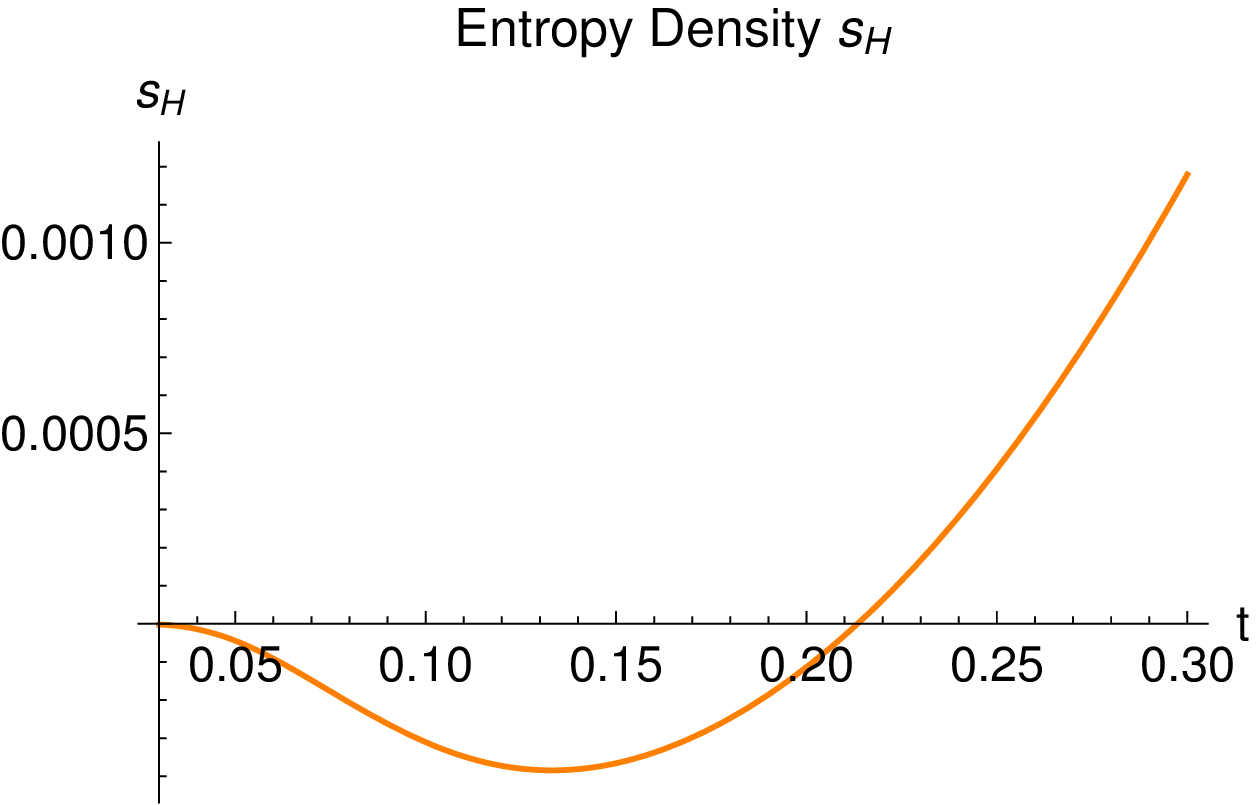}
\hspace{2mm}
\includegraphics[width=7.2cm]{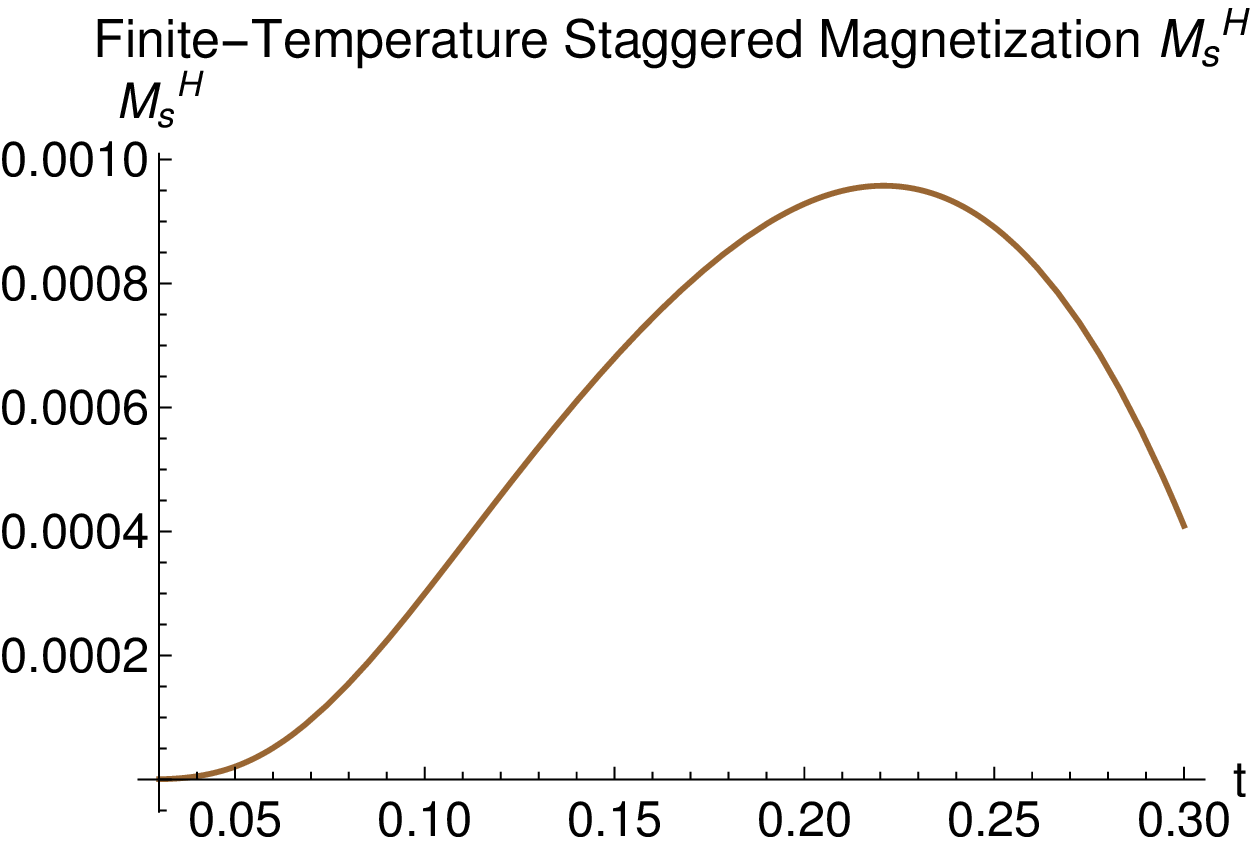}}
\vspace{4mm}
\hbox{
\includegraphics[width=7.2cm]{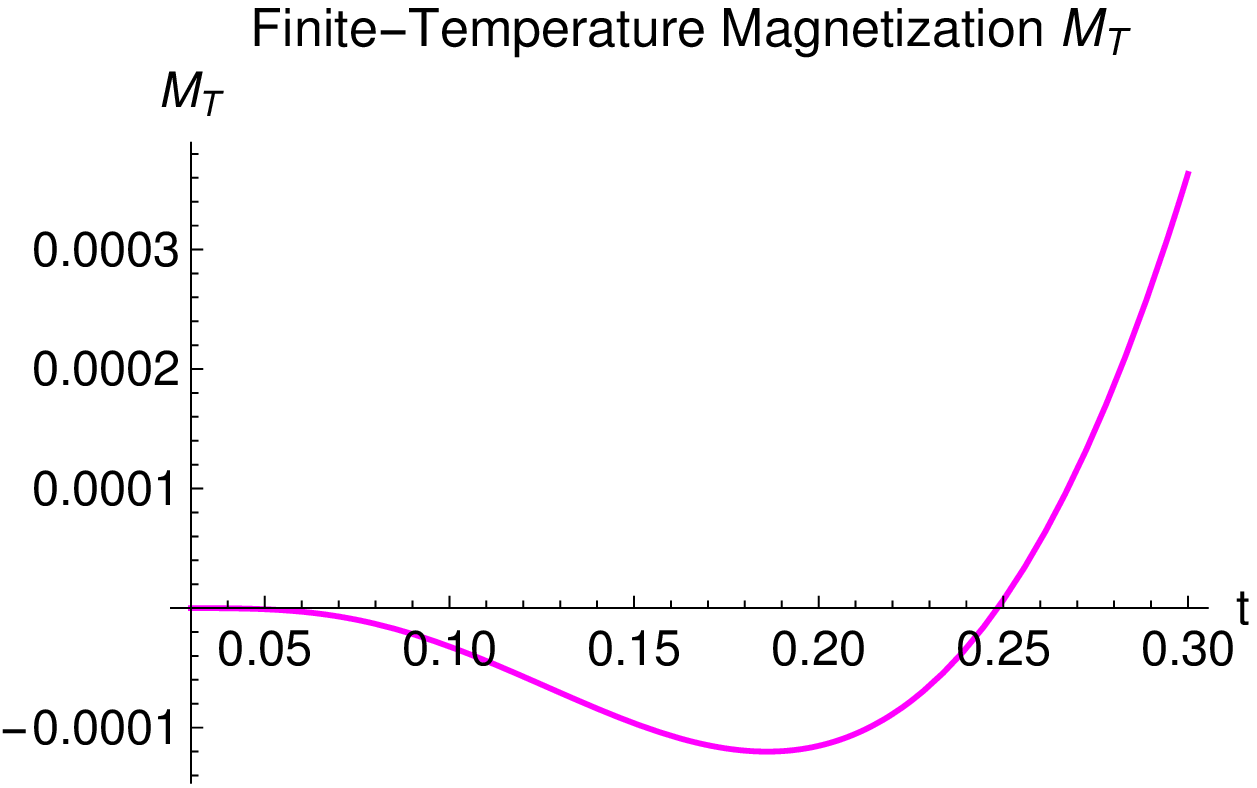}
\hspace{2mm}
\includegraphics[width=7.2cm]{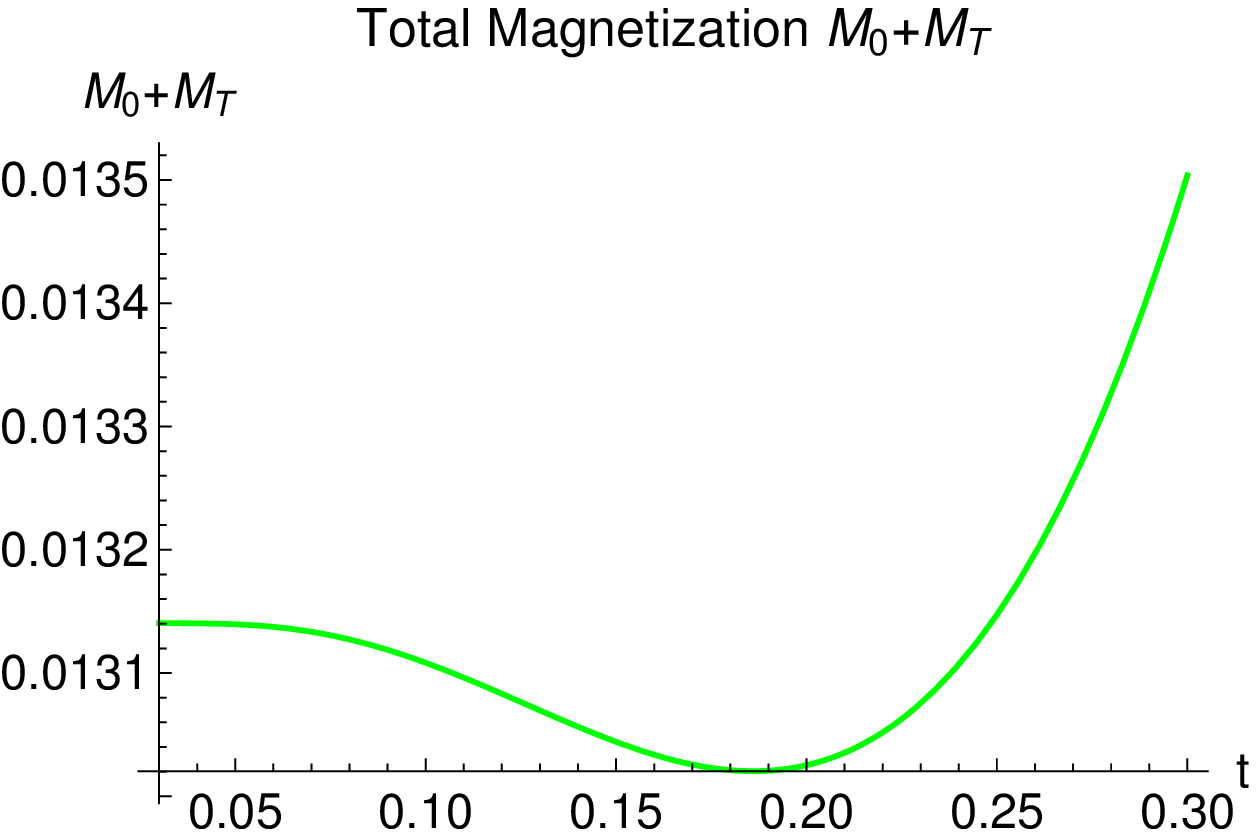}}
\end{center}
\caption{[Color online] Temperature dependence of the $H$-induced entropy density $s_H$, staggered magnetization $M^H_s$,  uniform
magnetization $M_T$, and total uniform magnetization $M_0 + M^T$ of the spin-$\frac{1}{2}$ square-lattice antiferromagnet in mutually
orthogonal staggered and magnetic fields of strength $(m,m_H)=(0.25,0.15)$.}
\label{figure7}
\end{figure}

Let us again reveal the impact of the magnetic field by considering the subtracted quantities,
\begin{eqnarray}
s_H & = & s(t,m,m_H) - s(t,m,0) \, , \nonumber \\
M^H_s & = & M^T_s(t,m,m_H) - M_s(t,m,0) \, ,
\end{eqnarray}
that measure the response of entropy density and finite-temperature staggered magnetization that is uniquely due to the magnetic field.
Here matters are quite different and even more intriguing than in the configuration of mutually aligned fields. While both quantities $s_H$
and $M^H_s$ also exhibit non-monotonic characteristics according to the upper panel of Fig.~\ref{figure7}, the entropy density $s_H$ first
drops, presents a minimum, and then starts to rise at more elevated temperatures. Analogously, the finite-temperature staggered
magnetization $M^H_s$ first rises, goes through a maximum and then falls off at more elevated temperatures. As before, the two quantities
are correlated, but the correlation of the extrema of $s_H$ and $M^H_s$ is not one-to-one: for the specific point $(m,m_H)=(0.25,0.15)$,
the entropy minimum occurs at $t=0.133$ while the staggered magnetization maximum is located at $t=0.221$. Still, an initial enforcement of
antiparallel spin order is detected both in $s_H$ and $M^H_s$. Note that the response of the system in mutually orthogonal fields is
exactly the opposite of what we observed in the case of mutually aligned fields: there, according to Fig.~\ref{figure2}, the magnetic field
initially destabilizes the antiparallel spin pattern and leads to a destruction of antiferromagnetic order: in mutually aligned fields,
initially $s_H$ increases and $M^H_s$ decreases. Notice also that these effects are more pronounced in the case of mutually aligned fields:
the extrema of $s_H$ and $M^H_s$ (cf. the scales in the respective horizontal axes of Fig.~\ref{figure2} versus Fig.~\ref{figure7}) differ
in about one order of magnitude.

We conclude that antiferromagnetic systems subjected to mutually orthogonal magnetic and staggered fields are more robust against
perturbations caused by temperature and the fields. If the fields are aligned, we have a conflicting situation: the staggered field forces
the spins to antialign in its own direction, but the magnetic field wants the spins to antialign in a plane transverse to it -- these two
tendencies clearly compete such that antiferromagnetically ordered spins exposed to mutually aligned fields are perturbed more drastically
by thermal fluctuations.

The behavior of the finite-temperature uniform magnetization $M_T$,
\begin{equation}
M_T = M(t,m,m_H)- M(0,m,m_H) \, ,
\end{equation}
that is oriented perpendicular to the staggered magnetization direction, is also quite remarkable. According to the lower panel of
Fig.~\ref{figure7} referring to the specific point $(m,m_H)=(0.25,0.15)$, $M_T$ first decreases at low temperatures, goes through a minimum,
then rises and eventually becomes positive. This is in stark contrast to the behavior in aligned fields where $M_T$ grows monotonically as
temperature rises (see Fig.~\ref{figure2}). It should be pointed out that the total uniform magnetization, i.e., the superposition of the
zero-temperature and finite-temperature portions, $M_0 + M_T$ (where $M_0 = M(0,0,m_H)$), always is positive as we illustrate in the lower
right of Fig.~\ref{figure7}. Negative values of $M_T$ are in fact expected and can be interpreted as thermal perturbations of the spins
that are tilted into the direction of the external magnetic field. What is really intriguing is that this perturbation gets weaker such
that $M_T$ presents a minimum, and that $M_T$ even becomes positive at more elevated temperatures: here, as in mutually aligned fields, the
creation of a uniform magnetization is enforced.

\begin{figure}
\begin{center}
\hbox{
\includegraphics[width=7.5cm]{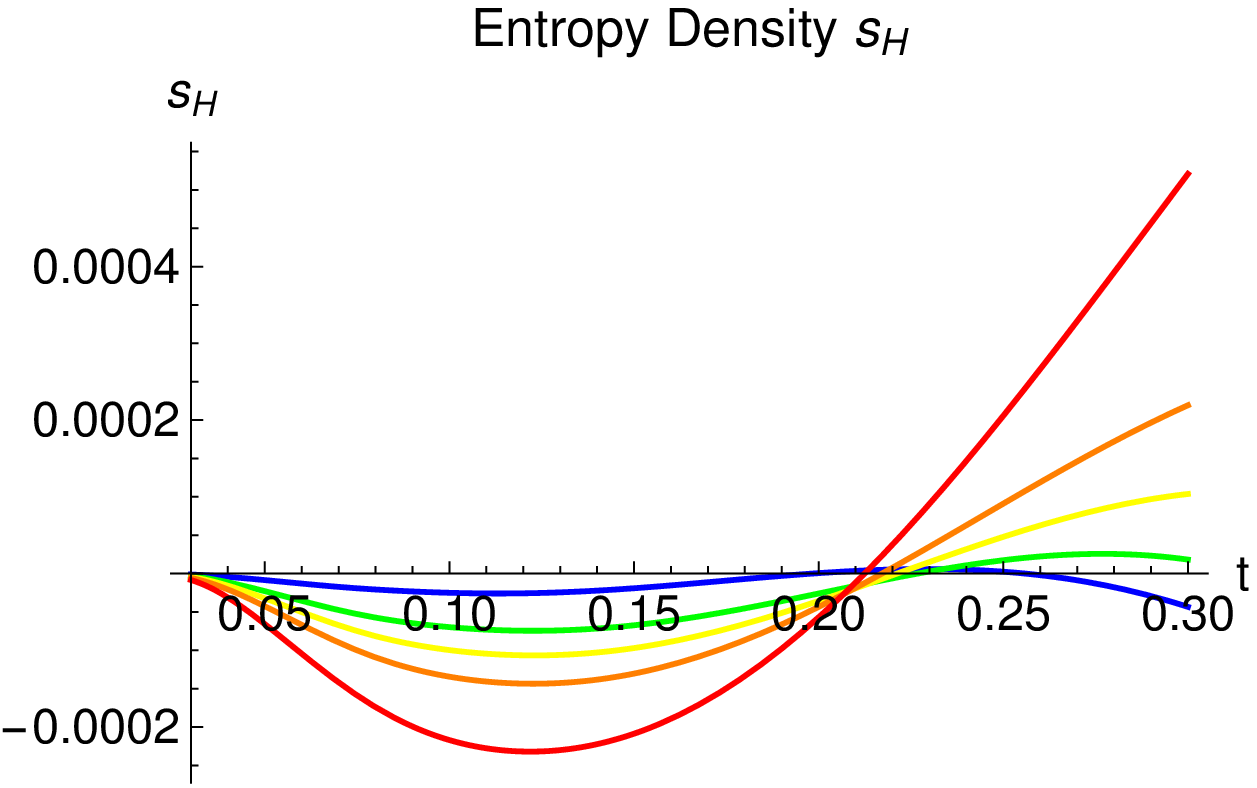}
\hspace{2mm}
\includegraphics[width=7.5cm]{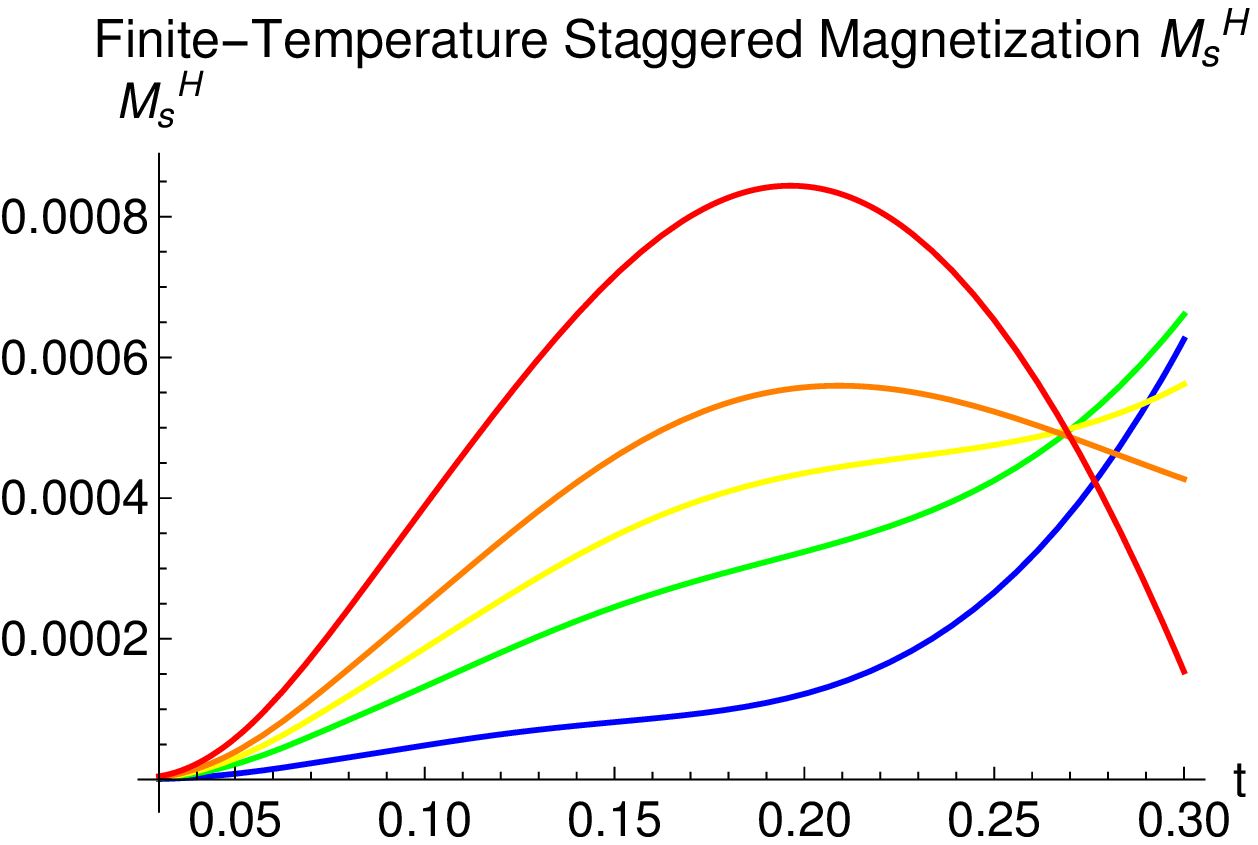}}
\end{center}
\caption{[Color online] Temperature dependence of the $H$-induced entropy density $s_H$ and staggered magnetization $M^H_s$ of the
spin-$\frac{1}{2}$ square-lattice antiferromagnet in mutually orthogonal staggered and magnetic fields of strength $m=0.18$ and
$m_H= \{ 0.03, 0.05, 0.06, 0.07, 0.09  \}$ (color-coded: blue to red).}
\label{figure8}
\end{figure}

In the mutually orthogonal case, things are still more complex as described so far. While in magnetic fields $m_H \gtrapprox 0.07$ the
system overall behaves as illustrated by the representative point $(m,m_H)=(0.25,0.15)$, in weak magnetic fields we observe in fact
a qualitatively different behavior both in the entropy density $s_H$ and the finite-temperature staggered magnetization $M^H_s$. This is
shown in Fig.~\ref{figure8} for points with staggered field strength fixed at $m=0.18$, but varying magnetic field strength
$m_H = \{ 0.03, 0.05, 0.06, 0.07, 0.09 \}$ -- color-coded from blue to red. The minimum-maximum characteristics of the curves for $s_H$ are
nicely reflected in the curves for $M^H_s$. Interestingly, in magnetic fields weak compared to the staggered field -- here
$m_H = \{ 0.03, 0.05, 0.06 \}$ smaller than $m=0.18 $ -- the finite-temperature staggered magnetization grows in the entire temperature
interval $0 < t < 0.3$: antiparallel alignment of the spins is enforced as temperature rises -- no maximum is present beyond which the
ordering effect is damped.

\begin{figure}
\begin{center}
\includegraphics[width=7.8cm]{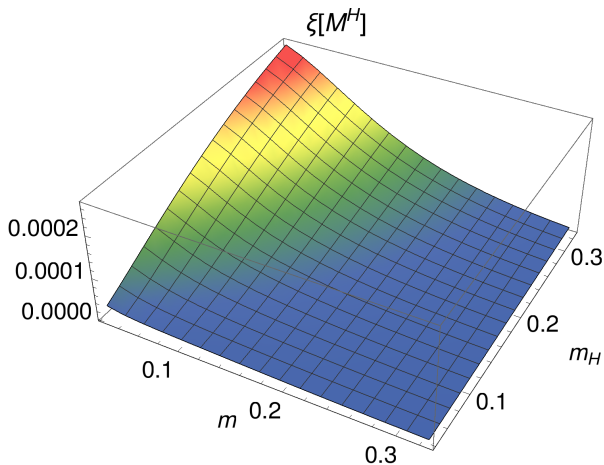}
\end{center}
\caption{[Color online] The parameter $\xi[M^H]$ for the spin-$\frac{1}{2}$ square-lattice antiferromagnet subjected to mutually orthogonal
magnetic ($m_H$) and staggered ($m$) fields.}
\label{figure9}
\end{figure}

Finally, the behavior of the finite-temperature uniform magnetization $M_T$ is qualitatively the same in the entire parameter region as for
the point $(m,m_H)=(0.25,0.15)$ referred to in Fig.~\ref{figure7}. In particular, also in weak magnetic fields, $M_T$ first drops, goes
through a minimum, rises and eventually becomes positive as temperature increases. To assess the phenomenon quantitatively, we scan the
surface $(m,m_H)$ and measure the parameter
\begin{equation}
\xi[M^H] = \int_0^{t_{\text{zero}}[M]} \!\!\! dt \, |M_T| \, ,
\end{equation}
where $t_{\text{zero}}[M]$ is the root of $M_T$.\footnote{For the point $(m,m_H)=(0.25,0.15)$ in Fig.~\ref{figure7} we have
$t_{\text{zero}}[M] \approx 0.25$.} The outcome is shown in Fig.~\ref{figure9}. One observes that in the region $m_H < m$ where we can compare
with the analogous parameter $\eta[M^H]$ for the mutually aligned configuration (see Fig.~\ref{figure5}), the induced uniform magnetization
is much smaller here: for the point $(m,m_H)=(0.25,0.15)$, e.g., we have $\xi[M^H] \approx 0.000013$ versus $\eta[M^H] \approx 0.00020$. On
the other hand, in the region $m_H > m$ that was not accessible due to the stability criterion before, in the present case of mutually
orthogonal fields the induced uniform magnetization $M_T$ is of the same magnitude as $\eta[M^H]$. The effect is more pronounced in weak
staggered fields ${\vec H}_s$. Although the limit ${\vec H}_s \to 0$ cannot be taken since the effective theory would break down,
extrapolating our results to very weak staggered fields correctly reproduces the behavior of the uniform magnetization that has been
reported in Refs.~\citep{FKLM92,San99,SS02,CRVV03,PR15} where the staggered field is absent: $M_T$ is negative and the effect is larger in
stronger magnetic fields.

\section{Conclusions}
\label{conclusions}

On the basis of magnon effective field theory we have performed a detailed survey of the thermomagnetic properties of bipartite
two-dimensional antiferromagnets exposed to magnetic and staggered fields. We have assessed the role of the magnetic field in
counterintuitive order-disorder phenomena that are evidenced by an increase of the uniform magnetization with temperature and the
non-monotonic temperature dependence of entropy density and staggered magnetization.

In the case of mutually aligned fields, the finite-temperature uniform magnetization grows monotonically. The phenomenon has so far not
been reported in theoretical studies -- with the exception of the effective field theory based Refs.~\citep{Hof20a,Hof21a}. Experimentally
it has been observed a long time ago in Ref.~\citep{AUW77}. In the case of mutually orthogonal magnetic and staggered fields, an even more
intriguing pattern emerges: the finite-temperature uniform magnetization first drops, goes through a minimum, starts to rise and eventually
tends to positive values at more elevated temperatures. Such intriguing behavior of the S=$\frac{1}{2}$ square-lattice isotropic
antiferromagnet so far has only been reported in studies where the staggered field was absent
(Refs.~\citep{FKLM92,San99,SS02,CRVV03,PR15,Iai18}).

We then have analyzed the behavior of entropy density and finite-temperature staggered magnetization. We observed that the entropy density
shift $s_H$ and the staggered magnetization shift $M_s^H$ caused by the magnetic field, are correlated and simultaneously monitor the
creation or destruction of antiparallel spin order. To reveal the existence of this correlation that is new to the best of our knowledge,
it was essential to first unmask the impact of the magnetic field by subtracting the $H$=0 portions in the entropy density and the
finite-temperature staggered magnetization, i.e., to define the quantities $s_H$ and $M_s^H$. The correlation between the extrema in the
temperature curves for $s_H$ and $M_s^H$ is almost one-to-one in the configuration of mutually aligned fields.

Interestingly, the respective response of antiferromagnetic monolayers subjected to mutually orthogonal fields is just the opposite of the
response that we observe in mutually aligned fields. In the latter case, the magnetic field initially destabilizes the antiparallel spin
arrangement that goes along with destruction of antiferromagnetic order: in mutually aligned fields, initially $s_H$ increases and $M^H_s$
decreases. In mutually orthogonal fields, on the other hand, antiparallel spin order is enhanced in presence of a magnetic field: here,
initially $s_H$ decreases and $M^H_s$ increases. Generically, the perturbation of the antiparallel spin pattern is more drastic in mutually
aligned fields. The fact that antiferromagnetic monolayers in mutually orthogonal fields are more robust against perturbations caused by
the fields and temperature, can be understood on the basis of conflicting tendencies imposed by the magnetic and the staggered field when
they are aligned.

Regarding the mechanism behind the observed phenomena, in the configuration where the fields are mutually orthogonal, the magnetic field
induces an easy plane and Kosterlitz-Thouless behavior emerges. Concretely, the minimum that we observe in the temperature dependence of the
uniform magnetization, reveals the presence of an easy plane. In contrast, in the configuration where magnetic and staggered fields are
mutually aligned, there is no "easy plane" induced by the magnetic field, but an "easy axis" induced by the staggered field. As a
consequence, XY behavior is absent and we observe monotonic dependence of the uniform magnetization with temperature. Non-monotonic
temperature dependence of the staggered magnetization and entropy density, however, emerges in either configuration of magnetic and
staggered fields. In mutually orthogonal fields, the finite-temperature staggered magnetization presents a maximum. Naively, the magnetic
field forces the spins to lie in the XY plane which enhances antialignment in the plane, whereas out-of-plane spin canting, i.e., the
creation of a uniform magnetization, is suppressed. In mutually aligned fields, the tendency of the magnetic field to destabilize the
antiferromagnetic spin arrangement and to align the spins in its proper direction dominates over thermal fluctuations, which causes the
uniform magnetization to increase. At the same time, antialignment of the spins along the same axis is reduced which leads to a decrease of
the finite-temperature staggered magnetization.

In all plots we refer to the spin-$\frac{1}{2}$ square-lattice antiferromagnet, but our rigorous and systematic two-loop analysis is valid
for any other two-dimensional bipartite lattice and for arbitrary spin. The various counterintuitive phenomena revealed by the effective
field theory investigation presented here are therefore universal.

As stated in the introductory section, the counterintuitive phenomena are not restricted to two spatial dimensions: they also arise in
three-dimensional antiferromagnets and in antiferromagnetic spin chains. Whereas the latter are not accessible within effective field
theory, three-dimensional systems are well within the scope of magnon effective field theory. Our study is moreover based on the isotropic
Heisenberg exchange model in a background of external magnetic and staggered fields. But one may envisage further types of interactions
such as spin-orbit coupling, Dzyaloshinskii-Moriya interaction -- to name but a few. It would be interesting to address the question of
emergence of magnetic order and non-monotonic behavior of entropy density, finite-temperature staggered and uniform magnetization also in
these more general settings. Respective work is in progress.

\section*{Acknowledgments}
The author gratefully acknowledges A.\ Iaizzi, A.\ W.\ Sandvik, and O.\ F.\ Syljuasen for correspondence.

\begin{appendix}

\section{Antiferromagnetic Monolayers in Mutually Parallel Magnetic and Staggered Fields}
\label{appendixA}

In this appendix we provide explicit two-loop effective field theory representations for the observables needed in our analysis: free
energy density, entropy density, staggered magnetization, and uniform magnetization. Readers interested in technical details of the
evaluation of the partition function for antiferromagnetic monolayers exposed to mutually aligned magnetic and staggered fields, are
referred to Ref.~\citep{Hof20a}. The free energy density amounts to
\begin{equation}
\label{freeEDtwoLoopParallel}
z = z_0 - {\hat g}_0 + \frac{H}{\rho_s} \, {\hat g}_1 \, \frac{\partial {\hat g}_0}{\partial H}
- \frac{\sqrt{M_s H_s} H}{4 \pi \rho_s^{3/2}} \, \frac{\partial {\hat g}_0}{\partial H}
- \frac{H^2}{\rho_s}{( {\hat g}_1)}^2
+ \frac{\sqrt{M_s H_s} H^2}{2 \pi \rho_s^{3/2}} \, {\hat g}_1 \, ,
\end{equation}
where the zero-temperature portion $z_0$ is
\begin{equation}
z_0 = - M_s H_s - \frac{M^{3/2}_s H^{3/2}_s}{6 \pi \rho_s^{3/2}} - (k_2 + k_3) \frac{M^2_s H^2_s}{\rho_s^2}
- \frac{M_s H_s  H^2}{16 \pi^2 \rho_s^2} \, .
\end{equation}
The finite-temperature piece ($z-z_0$) contains the dimensionful Bose functions ${\hat g}_r$, or, equivalently, the dimensionless Bose
functions ${\hat h}_r$,
\begin{eqnarray}
\label{g0g1g2Bose}
{\hat g}_0 \! & = &  \! T^3 \, {\int}_{\!\!\!0}^{\infty} \mbox{d} \lambda \, \lambda^{-5/2} e^{-\lambda m^2/4 \pi t^2}
\Bigg\{ \sqrt{\lambda} \, \theta_3\Big( \frac{m_H \lambda}{2 t}, e^{- \pi \lambda} \Big) e^{m_H^2 \lambda/4 \pi t^2} - 1 \Bigg\}
\equiv T^3 \, {\hat h}_0\, , \nonumber \\
{\hat g}_1  \! & = &  \! \frac{T}{4 \pi} \, {\int}_{\!\!\!0}^{\infty} \mbox{d} \lambda \, \lambda^{-3/2} e^{-\lambda m^2/4 \pi t^2}
\Bigg\{ \sqrt{\lambda} \, \theta_3\Big( \frac{m_H \lambda}{2 t}, e^{- \pi \lambda} \Big) e^{m_H^2 \lambda/4 \pi t^2} - 1 \Bigg\}
\equiv T \, {\hat h}_1 \, , \\
{\hat g}_2 \!  & = &  \! \frac{1}{16 \pi^2 T} \, {\int}_{\!\!\!0}^{\infty} \mbox{d} \lambda \, \lambda^{-1/2} e^{-\lambda m^2/4 \pi t^2}
\Bigg\{ \sqrt{\lambda} \, \theta_3\Big( \frac{m_H \lambda}{2t}, e^{- \pi \lambda} \Big)e^{m_H^2 \lambda/4 \pi t^2} - 1 \Bigg\}
\equiv \frac{{\hat h}_2}{T}  \, . \nonumber
\end{eqnarray}
The Jacobi theta function reads
\begin{equation}
\label{Jacobi3}
\theta_3(u,q) = 1 + 2 \sum_{n=1}^{\infty} q^{n^2} \cos(2 n u) \, ,
\end{equation}
and the three dimensionless and small parameters $m,m_H,t$ are defined in Eq.~(\ref{definitionRatios}).

The leading-order effective constants $M_s$ (staggered magnetization order parameter) and $\rho_s$ (spin stiffness), and the
next-to-leading order (NLO) effective constants $k_2$ and $k_3$, all depend on the geometry of the bipartite lattice. Our plots refer to
the spin-$\frac{1}{2}$ square-lattice antiferromagnet where the respective numerical values, following Ref.~\citep{GHJNW09}, are
\begin{eqnarray}
\label{squareLEC}
& & \rho_s = 0.1808(4) J \, , \quad M_s = 0.30743(1) / a^2 \, , \quad v = 1.6585(10) J a \, , \nonumber \\
& & \frac{k_2 + k_3}{v^2} = \frac{-0.0037}{2 \rho_s} = \frac{-0.0102}{J} \, .
\end{eqnarray}
Note that we also quote the result for the spin-wave velocity $v$ which is needed when we restore microscopic dimensions.

The two-loop effective field theory representation for the entropy density takes the form
\begin{equation}
s(t,m,m_H) = s_1 T^2 + s_2 T^3 + {\cal O}(T^4) \, ,
\end{equation}
with coefficients
\begin{eqnarray}
s_1 & = & \frac{t^2}{2} \, \frac{{\mbox{d}}^2 {\hat h}_{-1}}{\mbox{d} m_H^2}
+ \frac{m m_H t^2}{4} \, \frac{{\mbox{d}}^3 {\hat h}_{-1}}{\mbox{d} m_H^3}
- m_H (1 + \frac{m}{2}) \, \frac{\mbox{d} {\hat h}_0}{\mbox{d} m_H} \nonumber \\
& & - m m_H^2 \frac{{\mbox{d}}^2 {\hat h}_0}{\mbox{d} m_H^2}
+ \frac{m m_H^3}{t^2} \, \frac{\mbox{d} {\hat h}_1}{\mbox{d} m_H} \, , \nonumber \\
s_2 & = & - \frac{m_H t^2}{2 \rho_s} \, \frac{{\mbox{d}}^2 {\hat h}_0}{\mbox{d} m_H^2} \, \frac{\mbox{d} {\hat h}_0}{\mbox{d} m_H}
+ \frac{m^2_H}{\rho_s} \, \frac{\mbox{d} {\hat h}_0}{\mbox{d} m_H} \, \frac{\mbox{d} {\hat h}_1}{\mbox{d} m_H}
- \frac{m_H t^2}{2 \rho_s} \, {\hat h}_1 \, \frac{{\mbox{d}}^3 {\hat h}_{-1}}{\mbox{d} m_H^3} \nonumber \\
& & + \frac{2 m^2_H}{\rho_s} \, {\hat h}_1 \, \frac{{\mbox{d}}^2 {\hat h}_0}{\mbox{d} m_H^2}
- \frac{2 m^3_H}{t^2 \rho_s} \, {\hat h}_1 \, \frac{\mbox{d} {\hat h}_1}{\mbox{d} m_H}
+ \frac{m_H}{\rho_s} \, {\hat h}_1 \, \frac{\mbox{d} {\hat h}_0}{\mbox{d} m_H} \, .
\end{eqnarray}
The additional kinematical function ${\hat h}_{-1}$ is
\begin{equation}
\label{gMinus1Bose}
{\hat h}_{-1} = \frac{{\hat g}_{-1}}{T^5} = 4 \pi \, {\int}_{\!\!\!0}^{\infty} \mbox{d} \lambda \, \lambda^{-7/2} e^{-\lambda m^2/4 \pi t^2}
\Bigg\{ \sqrt{\lambda} \, \theta_3\Big( \frac{m_H \lambda}{2 t}, e^{- \pi \lambda} \Big) e^{m_H^2 \lambda/4 \pi t^2} - 1 \Bigg\} \, .
\end{equation}

The staggered magnetization amounts to
\begin{equation}
\label{OPAF}
M_s(t,m,m_H) = M_s(0,m,m_H) + {\tilde \sigma}_1 T + {\tilde \sigma}_2 T^2 + {\cal O}(T^3) \, ,
\end{equation}
where the coefficients are
\begin{eqnarray}
{\tilde \sigma}_1(t,m,m_H) & = & - \frac{M_s}{\rho_s} \, {\hat h}_1 \, , \nonumber \\
{\tilde \sigma}_2(t,m,m_H) & = & \frac{M_s}{\rho_s} \, \Bigg\{ \frac{m_H}{\rho_s} \, {\hat h}_2 \, \frac{\partial{\hat h}_0}{\partial m_H}
+ \frac{m_H}{\rho_s} \, {\hat h}_1 \, \frac{\partial{\hat h}_1}{\partial m_H}
+ \frac{m_H t}{8 \pi \rho_s m} \, \frac{\partial{\hat h}_0}{\partial m_H}  \nonumber \\
& & - \frac{m m_H}{4\pi \rho_s t} \, \frac{\partial{\hat h}_1}{\partial m_H}
- \frac{2 m_H^2}{\rho_s t^2} \, {\hat h}_1 {\hat h}_2
- \frac{m_H^2}{4\pi \rho_s m t} \, {\hat h}_1
+ \frac{m m_H^2}{2 \pi \rho_s t^3} \, {\hat h}_2 \Bigg\} \, .
\end{eqnarray}
Finally, for the uniform magnetization we have
\begin{equation}
\label{magnetizationAF}
M(t,m,m_H) = M(0,m,m_H) + {\hat \sigma}_1 T + {\hat \sigma}_2 T^2 + {\cal O}(T^3) \, ,
\end{equation}
with coefficients
\begin{eqnarray}
{\hat \sigma}_1(t,m,m_H) & = & 2 \pi \rho_s t^2 \frac{\partial {\hat h}_0}{\partial m_H} \, , \nonumber \\
{\hat \sigma}_2(t,m,m_H) & = & - 2 \pi t^2 {\hat h}_1 \frac{\partial {\hat h}_0}{\partial m_H}
- 2 \pi m_H t^2 \frac{\partial {\hat h}_1}{\partial m_H} \frac{\partial {\hat h}_0}{\partial m_H}
- 2 \pi m_H t^2 {\hat h}_1 \frac{\partial^2 {\hat h}_0}{\partial m^2_H} \nonumber \\
& & + \frac{m t}{2} \frac{\partial {\hat h}_0}{\partial m_H}
+ \frac{m m_H t}{2} \frac{\partial^2 {\hat h}_0}{\partial m^2_H}
+ 4 \pi m_H {({\hat h}_1)}^2
+ 4 \pi m^2_H {\hat h}_1 \frac{\partial {\hat h}_1}{\partial m_H} \nonumber \\
& & - \frac{2 m m_H}{t} {\hat h}_1
- \frac{m m_H^2}{t} \frac{\partial {\hat h}_1}{\partial m_H} \, .
\end{eqnarray}
Note that the staggered and uniform magnetization contain the zero-temperature portions
\begin{equation}
\label{OPT0}
\frac{M_s(0,m,m_H)}{M_s} = 1 + \frac{m}{2} + \frac{m_H^2}{4} + 8 \pi^2 \rho_s (k_2 + k_3) m^2 \, ,
\end{equation}
and
\begin{equation}
\label{MagT0}
M(0,m,m_H) = \pi \rho^2_s m^2 m_H \, ,
\end{equation}
respectively.

\section{Antiferromagnetic Monolayers in Mutually Orthogonal Magnetic and Staggered Fields}
\label{appendixB}

We now proceed along the same lines for antiferromagnetic monolayers that are subjected to mutually orthogonal magnetic and staggered
fields. Technical aspects on the evaluation of the partition function can be found in Refs.~\citep{Hof17,Hof21a}. The two-loop free energy
density amounts to
\begin{eqnarray}
\label{freeEDtwoLoopOrthogonal}
z & = & z_0 - \mbox{$ \frac{1}{2}$} \Big\{ g^{I}_0 + g^{I\!I}_0 \Big\} \nonumber \\
& & + \frac{M_s H_s}{16 \pi \rho^2_s} \, \Bigg\{ \sqrt{\frac{M_s H_s}{\rho_s} + H^2} - \sqrt{\frac{M_s H_s}{\rho_s}} \Bigg\} g^{I}_1
+ \frac{H^2}{4 \pi \rho_s} \, \sqrt{\frac{M_s H_s}{\rho_s} + H^2} \, g^{I}_1  \nonumber \\
& & - \frac{M_s H_s}{16 \pi \rho^2_s} \, \Bigg\{ \sqrt{\frac{M_s H_s}{\rho_s} + H^2} - \sqrt{\frac{M_s H_s}{\rho_s}} \Bigg\} g^{I\!I}_1 \\
& & - \frac{M_s H_s}{8 \rho^2_s} \, \Big\{ {(g^{I}_1)}^2 - 2 g^{I}_1 g^{I\!I}_1 + {(g^{I\!I}_1)}^2 \Big\}
- \frac{H^2}{2 \rho_s}{(g^{I}_1)}^2 + \frac{2}{\rho_s} \, s(\sigma,\sigma_H) \, T^4 \, , \nonumber
\end{eqnarray}
where the zero-temperature piece $z_0$ reads
\begin{eqnarray}
\label{vacuumEDtwoLoopOrthogonal}
z_0 & = & - M_s H_s - \mbox{$ \frac{1}{2}$} \rho_s H^2
- (k_2 + k_3) \frac{M^2_s H^2_s}{\rho^2_s} - k_1 \frac{M_s H_s}{\rho_s} H^2 -(e_1 + e_2) H^4 \nonumber \\
& & - \frac{1}{12 \pi} \Bigg\{ {\Big( \frac{M_s H_s}{\rho_s} + H^2 \Big) }^{3/2} + {\Big( \frac{M_s H_s}{\rho_s} \Big)}^{3/2} \Bigg\}
- \frac{M_s^2 H_s^2}{64 \pi^2 \rho^3_s} \nonumber \\
& & - \frac{5 M_s H_s H^2}{128 \pi^2 \rho^2_s} - \frac{H^4}{32 \pi^2 \rho_s}
+ \frac{M_s^{3/2} H_s^{3/2}}{64 \pi^2 \rho^{5/2}_s} \, \sqrt{\frac{M_s H_s}{\rho_s} + H^2} \, .
\end{eqnarray}
Note that $z_0$, apart from $k_2$ and $k_3$ that also arise in the case of mutually aligned fields, in addition involves the NLO effective
constants $k_1, e_1, e_2$.\footnote{The definition of the NLO effective constants $k_1,k_2,k_3,e_1,e_2$ is given in section 2 of
Ref.~\citep{Hof20a}.} The dimensionless Bose functions,
\begin{equation}
h^{I,{I\!I}}_0 = \frac{g^{I,{I\!I}}_0}{T^3} \, , \qquad h^{I,{I\!I}}_1 = \frac{g^{I,{I\!I}}_1}{T} \, , \qquad h^{I,{I\!I}}_2 = g^{I,{I\!I}}_1 \; T \, ,
\end{equation}
referring to magnon $I$ and magnon $I\!I$, respectively, are\footnote{Magnon $I$ and magnon $I\!I$ as defined by the dispersion relations
(\ref{disprelAFH}).}
\begin{eqnarray}
\label{BoseFunctions1}
h^{I}_0(H_s, H, T) & = & \frac{4 \pi^2 {(\sigma^2 + \sigma^2_H)}^{3/2}}{3}
- 2 \sqrt{\sigma^2 + \sigma^2_H} \; Li_2(e^{2 \pi \sqrt{\sigma^2 + \sigma^2_H} }) + \frac{1}{\pi} \; Li_3(e^{2 \pi \sqrt{\sigma^2 + \sigma^2_H} }) \nonumber \\
& & + 2 \pi (\sigma^2 +\sigma^2_H) \Big\{ \log(1- e^{-2 \pi \sqrt{\sigma^2 + \sigma^2_H}}) - \log(1- e^{2 \pi  \sqrt{\sigma^2 + \sigma^2_H} }) \Big\} \, ,
\nonumber \\
h^{I}_1(H_s, H, T) & = & - \frac{1}{2 \pi} \, \log \Big( 1 - e^{- 2 \pi\sqrt{\sigma^2 + \sigma^2_H}  } \Big) \, , \nonumber \\
h^{I}_2(H_s, H, T) & = & \frac{1}{8 \pi^2 \sqrt{\sigma^2 + \sigma^2_H}  \Big( e^{2 \pi \sqrt{\sigma^2 + \sigma^2_H} } - 1 \Big)} \, ,
\end{eqnarray}
and
\begin{eqnarray}
\label{BoseFunctions2}
h^{I\!I}_0(H_s, 0, T) & = & \frac{4 \pi^2 \sigma^3}{3} + 2 \pi \sigma^2  \Big\{ \log(1- e^{-2 \pi \sigma}) - \log(1- e^{2 \pi \sigma}) \Big\}
\nonumber \\
& & - 2 \sigma \; Li_2(e^{2 \pi \sigma}) + \frac{1}{\pi} \; Li_3(e^{2 \pi \sigma}) \, , \nonumber \\
h^{I\!I}_1(H_s, 0, T) & = & - \frac{1}{2 \pi} \, \log \Big( 1 - e^{- 2 \pi \sigma} \Big) \, , \nonumber \\
h^{I\!I}_2(H_s, 0, T) & = & \frac{1}{8 \pi^2 \sigma \Big( e^{2 \pi \sigma} - 1 \Big)} \, ,
\end{eqnarray}
where $Li_2, Li_3$ are polylogarithms. Here everything is given in terms of two dimensionless parameters $\sigma_H$ and $\sigma$ defined as
\begin{equation}
\label{defSigmas}
\sigma_H = \frac{H}{2 \pi T} \, , \qquad \sigma = \frac{\sqrt{M_s H_s}}{2 \pi \sqrt{\rho_s} T} \, .
\end{equation}
The connection between $\sigma_H, \sigma$ and $m_H, m$ is
\begin{equation}
\sigma_H = \frac{\rho_s}{T} \, m_H \, , \qquad \sigma = \frac{\rho_s}{T} \, m \, .
\end{equation}
The most complicated piece in the free energy density, Eq.~(\ref{freeEDtwoLoopOrthogonal}), is the dimensionless sunset function
$s(\sigma,\sigma_H)$ -- here we merely refer to Ref.~\citep{Hof17}, where the exact definition and a two-dimensional plot is provided by
Eq.~(B14) and Fig.~3, respectively.

The two-loop effective field theory representation for the entropy density takes the form
\begin{eqnarray}
\label{entropyTwoLoopOrthogonal}
s(t,m,m_H) & = & \frac{1}{2} \, \Bigg( \frac{\mbox{d} g^{I}_0}{\mbox{d}T} + \frac{\mbox{d} g^{I\!I}_0}{\mbox{d}T} \Bigg)
- \frac{\pi^2 \rho_s^2 m^2}{2} \, \Big( \sqrt{m^2 + m^2_H} - m \Big) \frac{\mbox{d} g^I_1}{\mbox{d}T} \nonumber \\
& & - 2 \pi^2 \rho_s^2 m^2_H \, \sqrt{m^2 + m^2_H}  \, \frac{\mbox{d} g^I_1}{\mbox{d}T}
+ \frac{\pi^2 \rho_s^2 m^2}{2} \, \Big( \sqrt{m^2 + m^2_H} - m \Big) \frac{\mbox{d} g^{I\!I}_1}{\mbox{d}T} \nonumber \\
& & + 2 \pi \rho_s m^2 \Bigg( g^I_1 \, \frac{\mbox{d} g^I_1}{\mbox{d}T}
+ g^{I\!I}_1  \, \frac{\mbox{d} g^{I\!I}_1}{\mbox{d}T}
- g^{I\!I}_1  \, \frac{\mbox{d} g^I_1}{\mbox{d}T}
- g^I_1  \, \frac{\mbox{d} g^{I\!I}_1}{\mbox{d}T} \Bigg) \nonumber \\
& & + 4 \pi^2 \rho_s m^2_H g^I_1  \, \frac{\mbox{d} g^I_1}{\mbox{d}T}
- \frac{2}{\rho_s} \frac{\mbox{d} s(\sigma,\sigma_H)}{\mbox{d}T} \, T^4
- \frac{8}{\rho_s} \, s(\sigma,\sigma_H) \, T^3 + {\cal O}(T^4) \, .
\end{eqnarray}

Finally, the low-temperature expansions for the staggered and uniform magnetization are
\begin{eqnarray}
\label{OPMagOrthogonal}
M_s(t,m,m_H) & = & M_s(0,m,m_H) + {\tilde \sigma}_1 T + {\tilde \sigma}_2 T^2 + {\cal O}(T^3) \, , \nonumber \\
M(t,m,m_H) & = & M(0,m,m_H) + {\hat \sigma}_1 T + {\hat \sigma}_2 T^2 + {\cal O}(T^3) \, ,
\end{eqnarray}
with coefficients
\begin{eqnarray}
{\tilde \sigma}_1(t,m,m_H) & = & -\frac{M_s}{2 \rho_s} \, \Big( h^{I}_1 + h^{I\!I}_1 \Big) \, , \nonumber \\
{\hat \sigma}_1(t,m,m_H) & = & - 2 \pi \rho_s m_H h^{I}_1 \, .
\end{eqnarray}
Since the explicit expressions for coefficients ${\tilde \sigma}_2$ and ${\hat \sigma}_2$ are quite lengthy, we do not list them here. Note
that their derivation -- although cumbersome -- is trivial via
\begin{eqnarray}
M_s(T,H_s,H) & = & - \frac{\partial z(T,H_s,H)}{\partial H_s} \, , \nonumber \\
M(T,H_s,H) & = & - \frac{\partial z(T,H_s,H)}{\partial H} \, .
\end{eqnarray}
The zero-temperature portions in the staggered and uniform magnetizations are (see Ref.~\citep{Hof17})
\begin{eqnarray}
\label{OPorthogonalT0}
\frac{M_s(0,m,m_H)}{M_s} & = & 1 + \frac{m}{4} + \frac{\sqrt{m^2+m_H^2}}{4} + \frac{m^2}{8} + \frac{5 m_H^2}{32}
- \frac{m^3}{8 \sqrt{m^2+m_H^2}} \nonumber \\
& & - \frac{3 m \, m_H^2}{32 \sqrt{m^2+m_H^2}} + 8 \pi^2 \rho_s (k_2 + k_3) \, m^2 + 4 \pi^2 \rho_s k_1 \, m_H^2 \, , \nonumber \\
& & \hspace{-2.4cm} m = \frac{\sqrt{M_s H_s}}{2 \pi \rho^{3/2}_s} \, , \qquad m_H = \frac{H}{2 \pi \rho_s} \, , \qquad 
M_s = M_s(0,0,0) \, ,
\end{eqnarray}
and
\begin{eqnarray}
\label{MagorthogonalT0}
\frac{M(0,m,m_H)}{\rho^2_s} & = & 2 \pi \, m_H + \pi \, m_H \sqrt{m^2+m_H^2} + \pi m_H^3 + \frac{5 \pi}{8} \, m^2 \, m_H \\
& &- \frac{\pi}{8} \, \frac{m^3 \, m_H}{\sqrt{m^2+m_H^2}} + 32 \pi^3 \rho_s (e_1 + e_2) \, m_H^3 + 16 \pi^3 \rho_s k_1 \, m^2 m_H \, ,
\nonumber
\end{eqnarray}
respectively.

\end{appendix}

\end{document}